\documentclass[12pt]{report}
\begin{document}
\textheight=22cm
\textwidth=17.4cm
\newcommand{\be}{\begin{equation}}
\newcommand{\Lam}{\Lambda}
\newcommand{\ee}{\end{equation}}
\newcommand{\lam}{\Lambda}
\newcommand{\gam}{\gamma}
\newcommand{\Gam}{\Gamma}
\newcommand{\bt}{\beta}
\newcommand{\alf}{\alpha}
\newcommand{\rh}{\rho}
\newcommand{\bc}{\begin{center}}
\newcommand{\ec}{\end{center}}
\newcommand{\fr}{\frac}
\newcommand{\dt}{\dot}
\newcommand{\dd}{\ddot}
\newcommand{\om}{\omega}
\newcommand{\se}{\section}
\baselineskip=20pt
\begin{center}
\Large{\bf{Nonstandard Cosmology With Constant and Variable
Gravitational and Variable Cosmological ``Constants" and Bulk Viscosity}}
\end{center}
\vspace{1.5cm}
\bc
 A thesis submitted for the degree of Doctor of Philosophy (Physics)
\ec
\vspace{0.5cm}
\bc
{\it by}
\ec
\bc
{\bf  Arbab Ibrahim Arbab }\\
\ec
\vspace{1cm}

{\it Supervised by}

{{\sf\bf Professor A-M. M. Abdel Rahman}}
\vspace{3cm}
\newline
{{\it {Department of Physics,\\
Faculty  of Science\ ,\\
University of Khartoum, P.O. Box 321,\\
Khartoum 11115 -- Sudan.}}\\
\vspace{1.5cm}
{\tt October, 1996.}}
\pagenumbering{roman}
\baselineskip=20pt
\small
\baselineskip=20pt
\tableofcontents
\newpage
\vspace{1in}
\bc
{\bf ACKNOWLEDGMENTS}
\ec
\small
\baselineskip=20pt
\textheight=9in
I would like to express my gratitude and debt to my supervisor, Prof. A-M. M. Abdel Rahman
for his generous care and his guidance during the entire work. His continuous
encouragement is dearly acknowledged. I am very grateful to his assistance and help.

I wish to thank my friends at the Department of Physics who
painstakingly helped me in typing the manuscript and for their
encouragement and endeavor. I would also like to thank the
International Centre for Theoretical Physics (ICTP) for their
hospitality at the Centre and the Kuwait Arab Fund for their kind
support to travel to the Centre. Special thanks go to Mr. Madi
Musa for his kind help in sending extracts from the manuscript for
publication. I extend my thanks to the Physics Department,
University of Khartoum, where this work was carried out.
\newpage
\vspace{1in}
\bc
{\bf ABSTRACT}
\ec
We have analyzed a nonsingular model with a variable cosmological term
following the Carvalho {\it et al}. ansatz. The model was shown to approximate to
 the model of Freese {\it et al}.
in one direction and to the \"{O}zer-Taha in the other. We have then included the
effect of viscosity in this cosmology, as this effect has not been considered
before. The analysis showed that this viscous effect could be important
with a present contribution to the cosmic pressure, at most, of order of
that of radiation. The model puts a stronger upper bound  on the baryonic matter
than that required by the standard model.
A variable gravitational and cosmological constant  were then
introduced in a scenario which conserves the energy and momentum in the
presence of bulk viscosity. The result of the analysis reveals that various
models  could be viscous. A noteworthy result is that some nonsingular closed models evolve
asymptotically into a singular viscous one. The considered models solve for many of
the standard model problems. Though the introduction of bulk viscosity results in
the creation of particles, this scenario conserves energy and momentum. As in the
standard model the entropy remains constant. We have not explained the generation
of bulk viscosity but some workers attributes this to neutrinos. Though the
role of viscosity today is minute it could, nevertheless, have had an important
contribution at early times. We have shown that these models encompass many
of the old and recently proposed models, in particular, Brans-Dicke, Dirac,
Freese {\it et al}., Berman, Abdel Rahman and Kalligas {\it et al}.
models. Hence we claim that the introduction of bulk viscosity enriches
the adopted cosmology.
\chapter{Introduction}
\pagenumbering{arabic}
The standard model was  successful in describing the Universe, since it predicts
the existence of the observed cosmic microwave background radiation having
a temperature of $2.75 \rm K$. The other success was the formation of light elements during
the first few minutes after the big bang. However, this model is fraught with some
vexing problems. These problems remain unsolved within the context of the standard model.
An extension to this model becomes inevitable, this extension may include an
extension to the theory of general relativity.

Einstein introduced a cosmological term to the field equations to obtain a
 static Universe, but later, following the discovery of cosmic expansion, discarded this term.
However, there are good reasons to believe that this term had an appreciable
 contribution at the beginning of the Universe. From a point of view of quantum
 field theory this term could correspond to a vacuum energy  which was found to
 be very much bigger than the observed value. This discrepancy is known as the cosmological
constant problem. In Particle Physics, the solution of this problem amounts to
finding a mechanism  that requires this term  to vanish.
An alternative to this view is to assume that as the Universe evolves, this term
evolves and decreases to its present value. According to this view the cosmological
term is small because the Universe is too old. Recently, cosmological models with
this term decreasing with time were proposed (\"{O}zer \& Taha, Freese {\it et al}.,
Gasperini, Chen \& Wu, Abdel Rahman, Carvalho {\it et al}. ). These models are known
as vacuum decaying models.

In Chapter 2 and 3, we review the standard model concentrating on
its successes and shortcomings. Chapter 4 deals with a general review to the
nonstandard cosmology. A nonsingular model depending on the Carvalho {\it et al}.
ansatz ($\Lambda=3\bt H^2+\fr{3\alf}{R^2}$, where $\alpha, \beta$ are arbitrary
dimensionless constants) is introduced in Chapter 5.
In Chapter 6 we present a closed nonsingular viscous model
with a cosmological constant of the Chen and Wu type and discuss its implications
for the evolution of the Universe.  Chapter 7 is devoted to a singular
model with  varying cosmological and gravitational constants containing
bulk viscosity varying as $\rh^n$, where $\rh$ is the density of matter
and $n$ is a positive number.
Various models are reproduced with appropriate values for  $n$.
A flat viscous model with $\beta=0$ is presented in Chapter 8.
The result of the analysis shows that $G$ increases with time in both radiation
dominated and matter dominated phases of cosmic evolution.
These results are in accordance with the previously found results of chapter 7.
An appendix  followed by some published papers is presented at the end of the thesis.

Finally we remark that in these cosmologies we have not discussed the issue of influence of bulk
viscosity on the  formation of galaxies. We will treat this in a future work.
\chapter{ The Standard Model}
\se{Background} A cosmological model is a model of our Universe,
taking into account and using all known physical laws, predicts
(approximately) correctly the observed properties of the Universe,
and in particular explains in detail the phenomena in the early
Universe. In a more restricted sense cosmological models are
solutions of Einstein's field equations for a perfect fluid that
reproduce the important features of our Universe. All cosmological
models which differ only  near the origin of the Universe must be
accepted as equally valid. In fact a series of solutions are known
which are initially inhomogeneous or anisotropic to a high degree,
and which then increasingly come to approximate the observed
Universe. All cosmological models which yield a red-shift($z$) and
a cosmic background radiation can hardly be refuted. The
possibility  cannot be excluded that our Universe is neither
homogeneous  nor isotropic, but has those properties only
approximately in our neighborhood. Every model is of course a
great simplification of reality, and only by the study of many
solutions can one establish which simplifications are allowed and
which assumptions are essential. \se{Einstein's field equations}
These equations describe the gravitational field resulting from
the distribution of matter  in the Universe. They are nonlinear
partial differential equations.
\be
{\cal{R}}_{\mu\nu}- \fr{1}{2} {\cal {R}}g_{\mu\nu}=8\pi G T_{\mu\nu}
\ee
where
$R_{\mu\nu} $ is the Ricci tensor and $\cal{R}$ is the curvature scalar, $T_{\mu\nu}$
is the energy-momentum tensor of the source producing the gravitational
field, and $G$ is Newton's gravitational constant.\\
The energy momentum tensor satisfies the following  requirements.\\
(i)  $  T_{\mu\nu}$ is symmetric with respect to interchange of $\mu$ and $\nu$ .\\
(ii)  $ T_{\mu\nu}$ is divergenceless, for energy and momentum to be conserved,\\
(Bianchi identity):
\be
T^{\mu\nu} ;\nu = 0
\ee
The Ricci tensor is defined by
\be
{\cal{R}}_{\mu\nu}=\fr{\partial^2 \ln\sqrt{-g}}{\partial x^{\mu}\partial x^{\nu}}-\fr{\partial \Gam^{\ell}_{\mu\nu}}{\partial x^{\ell}}+\Gam^m_{\mu n}\Gam^n_{\nu m}-\Gam^{\ell}_{\mu\nu}\fr{\partial \ln\sqrt{-g}}{\partial x^{\ell}}\ ,
\ee
and
\be
{\cal{R}} = g^{\mu\nu} R_{\mu\nu}\ ,
\ee
where $ g_{\mu\nu} $ is the metric tensor, $g$ its determinant and $ \Gam^{i}_{jk}$
are the Christoffel symbols related to  $ g_{\mu\nu}$ by
\be
\Gam^ {i}_{j k} = g^{i\alf} (g_{\alf j,k} + g_{\alf k,j} - g_{j
k,\alf}). \ee \se{Robertson-Walker Metric} The distribution of
matter and radiation in the observable Universe is homogeneous and
isotropic. While this by no means guarantees that the entire
Universe is smooth, it does  imply that  a region at least as
large as our  present observable Universe (Hubble's volume) is
smooth. So long as the Universe is spatially homogeneous and
isotropic on a scale as large as Hubble's volume, for purposes of
our local Hubble's volume we may assume that the entire Universe
is homogeneous and isotropic. This is known as {\it the
Cosmological Principle}. The metric for the space  with
homogeneous and isotropic sections which being maximally
symmetric is the Robertson - Walker (RW),
\be
ds^2 =dt^2 -R^2(t)(\fr{dr^2}{1-kr^2} +r^2(d\phi^2 +\sin ^2\theta d^2\phi))
\ee
where $(t,r,\theta,\phi)$ are coordinates (referred to as comoving coordinates), $R(t)$ is the cosmic scale factor,
and with appropriate scaling of the coordinates, $k$ can be chosen  to be  1, -1, or 0
for spaces of constant positive (closed), negative (open), or spatial (flat) curvature, respectively.
The time coordinate in the above equation  is the proper (or clock) time- measured by  an observer
at rest in the comoving frame, i.e. $(r,\theta,\phi) = \rm constant $
\se{Expansion of the Universe}
The  RW line element contains the function $R(t)$, which can be any function of time, t.
This line element  necessarily requires that the Universe cannot be static.
Because of the presence  of the term  $R(t)$, an element of the spatial distance $dl$ changes
with time, i.e. the distance between any pair of galaxies changes with time.
 Because of homogeneity, the cosmological fluid can not sustain any pressure - gradient,
the concomitant non-gravitational forces are absent. But when the only force present is
 gravitation, a static Universe is evidently not possible, it must either contract under
 gravity or expand against gravity.
 However, the analysis of the observed data from distant galaxies shows that the Universe
  is expanding. The recession velocity of a galaxy turns out to be related to
  its distance ($d$) by a simple law
$  v = H d$  where the proportionality constant $H$ is the Hubble constant.
  Note that the recession has no effect whatsoever on individual bodies since
  a homogeneous medium generates no gravitational field inside a spherical cavity.
\se{Distance measurements in RW metric}
The determination of distance in astronomy is mostly done using the concepts
and ideas of a three-dimensional Euclidean space. We therefore want to describe
briefly how the laws of light propagation in RW metrics influence this distance.
One possible way of determining the distance of an object is to compare its absolute
brightness $L$, which is defined as the total radiated energy per unit time and is
regarded as known, with the apparent brightness $I$ of the energy reaching the receiver
per unit time per unit surface area. The {\it Luminosity Distance } $D_L$ is defined by
\be
D_L=\sqrt{L/4\pi I}
\ee
so that in Euclidean space luminosity distance and geometrical distance coincide.
eq.(2.6) can be written as
\be
ds^2=dt^2-R^2[d\chi^2+f(\chi)(d\theta^2+\sin\theta^2d\phi^2)]
\ee
where
\be
\chi=\fr{r}{\sqrt{1-kr^2}}
\ee
In the RW space-time there exists a complicated relationship between the
true distance $D$ and the brightness distance $D_L$.
Due to the red-shift of the emitted light this becomes [5],
\be
D_L=(1+z)D\fr{f(\chi)}{\chi}, \ \ \
\ee
where
$$\left\{        f(\chi)
\begin{array}{c} =\sin\chi \ \ , k=1 \\
                 =\chi \ \ \ , k=0 \\
                 =\sinh\chi \ \  \ , k=-1
\end{array}
\right\}$$
Since one observes stars with $z\gg 0$, $D$ and $D_L$ can differ from one another
considerably.

A second possible way of determining distance is to compare the true diameter $\Delta$
of a cosmic source with the angle $\delta$ which it subtends at the Earth.
In the RW metric, eq.(2.8) implies,
\be
D_A\equiv\fr{\Delta}{\delta}=f(\chi)R(t_1)=\fr{D}{1+z}\fr{f(\chi)}{\chi}
\ee
These two examples of how to determine distance show clearly how the space
curvature comes into astronomical considerations concerning the law of
propagation of light. Of course optical methods can only be used to determine
the distances of objects whose light reaches us.
\se{The Friedmann equations}
  Having establish the metric (RW), the solution of Einstein's field
  equations requires a knowledge of the form of the energy-momentum tensor,
 $T_{\mu\nu}$. To be consistent with the symmetry of the metric we find that
 the simplest realization of such a tensor, $T_{\mu\nu}$, is that of a
 perfect fluid characterized by a time-dependent energy density $\rh(t)$ and
  pressure $p(t)$. For a perfect fluid symmetry requirements dictate that the
  energy momentum tensor has the form:
  \be
  T_{\mu\nu}=(\rh+p)u_\mu u_\nu -pg_{\mu\nu}\ ,
  \ee
  where  $u^\mu $ is the  fluid four-velocity and for a comoving frame,
 $u^\mu=(1,0,0,0) $ ( i.e. a fluid at rest).

The equation of motion of a particle in a gravitational field is given by
    \be
  \fr{du^\mu}{ds}+\Gam^\mu_{\Lambda\nu}u^\Lambda u^\nu=0
  \ee

  Straightforward but tedious calculations show that the components of the
  Ricci tensor are [5]
  \be
  {\cal{R}}_{00}=-3\fr{\dd R}{R},
  \ee
  \be
{\cal{R}}_{ij}=-(\fr{\dd R}{R}+\fr{2\dt R^2 +2k}{R^2})g_{ij},
  \ee
  and the curvature scalar is
  \be
{\cal{R}}=-6(\fr{\dd R}{R}+\fr{\dt R^2 +k}{R^2}).
\ee
The spatial 3-dimensional curvature is given by
\be
^3{\cal{R}}=\fr{6k}{R^2}
\ee
From eqs.(2.12) and (2.14-16), it follows that
\be
\fr{\dt R^2}{R^2}+\fr{k}{R^2}=\fr{8\pi G}{3}\rh,
\ee
 and
 \be
2\fr{\dd R}{R}+\fr{\dt R^2}{R^2}+\fr{k}{R^2}=-8\pi G p.
 \ee
 The Einstein's field equations (EFE) are related by the Bianchi
 identities and only two are independent. From eqs.(2.18) and (2.19) we obtain
 \be
 \fr{\dd R}{R}= -\fr{4\pi}{3}G(\rh +3p)
 \ee
Today $\dt R \ge 0$; if in the past $p+3\rh$ was always positive, then
$\dd R$ was always negative, and thus at finite time in the past
$R$ must have been equal to zero. This event, referred to as the {\it big bang},
is usually taken at time $t=0$. \\ At  $R=0$, there is a singularity; extrapolating
past the singularity is not possible in the framework of classical general
relativity.\\
The constant defined by $H=\fr{\dt R}{R}$  is not a genuine constant but in  general
varies with time as $t^{-1}$  and hence $H^{-1}$ sets  a time scale for the expansion :
$R$ roughly doubles in a Hubble time.

Equations (2.18) and (2.19) are known as Friedmann's equations.
eq.(2.18)  can be written as
\be
\fr{k}{H^2R^2}=\fr{\rh}{3H^2/8\pi G}-1\equiv\Omega-1
\ee
where
$\Omega=\rh/\rh_c $ and $\rh_c=3H^2/8\pi G $:\ \ \
$\rh_c $ is known as {\it the critical density} of the Universe and $\Omega$
as {\it the density parameter}. eqs.(2.18) and (2.20) give
\be
\fr{k}{R^2}=(2q-1) H^2
\ee
where $q=-R\dd R/\dt R^2$, a dimensionless parameter, is known as {\it the deceleration parameter}
which is a measure of slowing down the expansion of the Universe, is
\be
q=\fr{\Omega}{2}(1+3p/\rh)
\ee
Hence $q_0=\fr{\Omega_0}{2}(1+3p_0/\rh_0) $, where the subscript 0 denotes the
present day quantities.
The sign of $k$ in eqs.(2.21) and (2.22) gives the sign of $\Omega-1$ and $q$.
At the moment observations only yield the bound $-1< q_0 < 2$.

The parameter $k$ is normalized to the values
$$ k=+1 \ \ \rightarrow \Omega > 1 \ \ closed $$
$$ k=0 \ \ \rightarrow \Omega = 1 \ \ flat $$
$$ k=-1 \ \ \rightarrow \Omega < 1 \ \ open $$
We will consider each of the above cases separately.

\se{Successes and shortcomings of the Standard Model}
The standard big-bang model of the Universe had three major successes:\\
(i) it predicts that something  like Hubble's law must hold for the Universe.\\
(ii)  it predicts successfully the formation of light atomic nuclei from
protons and neutrons a few minutes after the big-bang. This prediction gives
the correct abundance ratios for $^3He, D, ^4He $ and $^7Li$ .\\
(iii) it predicts a relic of cosmic background radiation having a black-body spectrum
with a temperature of $2.75 \rm K,$ today .\\
However, certain problems and puzzles remain in the standard model.\\
(i) the Universe displays a remarkable degree of large-scale homogeneity.
This is evident in the cosmic microwave background radiation (CMBR)
which is known to be uniform  in temperature  to  about one part in $10^4$.\\
(ii) a certain amount of inhomogeneity must have existed in the primordial
matter to account for the clumping of matter in galaxies and cluster of
galaxies, etc., that we observe today. Any small inhomogeneity in the primordial
matter rapidly  grows into a large one with gravitational self-interaction. Thus
one has to assume a considerable smoothness in the primordial matter to account
for the inhomogeneity in the scale of galaxies at the present time.
The problem becomes acute if one extrapolates to $10^{-43}$s after the big-bang
when one has to assume an unusual smoothness in the initial state of matter.
This is known as the {\it smoothness problem}.\\
(iii) the present discrepancy between the observed density of matter and the required value. If $\Omega$ were initially equal
to unity ( flat Universe) it will stay equal to unity forever.
 On the other hand, if $\Omega$ were initially different from unity, its
 departure from unity will increase with time.\\
 The present value of $\Omega$ ranges between 0.1 and 2. For this to be the case
 the value of $\Omega$ would have had to be equal to 1 to one part in $10^{15}$
 a second or so after the big-bang, which seems an unlikely situation.
 This is called the {\it flatness problem}.\\
 To deal with these problems Alan Guth (1981), proposed  a model of the Universe,
 {\it known as the inflationary model}, which does not differ from the standard
 model  after a fraction of a second or so, but from about $10^{-45}$ to $10^{-30}$ seconds
  it has a period of extraordinary expansion, or {\it inflation}, during which
  a typical distance ($R$) increases by a factor of about $10^{50}$ more than the
  increase that would obtain in the standard model. Though inflationary
  models solve some of the problems of the standard model, they throw up problems
  of their own, which have not all been dealt with in a satisfactory manner [6].\\
  The consideration of the Universe in the first second or so calls for a great deal of
  information from the theory of elementary particles, particularly in the inflationary
  models. This period is referred to as {\it the very early Universe}.
\se{Matter dominated (MD) Universe}
{1.\it Flat model}\\
This is the simplest case ($k=0$) and is known as the Einstein de Sitter (ES) model.
Equations (2.2) and (2.12) give
\be
\dt\rh +3\fr{\dt R}{R}(\rh+p)=0\ \ .
\ee
The pressure $p$ and the density $\rh$ are related by the equation of state:
\be
p=p(\rh)
\ee
which is taken as $p=w \rh  $ for a perfect fluid.\\
For dust $w$ = 0, for radiation $w$= 1/3 and $w $= -1 for a vacuum
dominated Universe.
Hence
\be
\rh\sim R^{-3(1+w)}
\ee
For the matter dominated epoch $p=0$, i.e. $w$=0, therefore
eq.(2.24) gives
\be
R=R_0(t/t_0)^{2/3}
\ee
where
\be
\rh_0R^3_0=\rh R^3\ .
\ee
For this case $\rh=\rh_c , \ \Omega=1 $. The present value of $\rh_c$ is
$2\times 10^{-29}\rm gcm^{-3}$. The present age of the Universe, $t_0$ is given by
\be t_0 =\fr{2}{3H_0}
\ee
which has been found to be less than the range allowed by observations.\\
2.{\it closed model}\\
This is the case  $k = +1$. eqs.(2.20) and (2.18) give
$$ \fr{\dd R}{R}=-\fr{4\pi}{3}G\rh $$
and
$$ \fr{\dt R^2}{R^2}=\fr{8\pi G}{3}\rh-\fr{1}{R^2} $$
The age of the Universe, $t_0$, is given from the integral
\be
t=\int\fr{\sqrt{R}dR}{\sqrt{\alf-R}}
\ee
where $\alf=\fr{2q_0}{H_0(2q_0-1)^{3/2}}$\ \ \ .
One has
\be
t_0=\fr{q_0}{H_0(2q_0-1)^{3/2}}[\cos^{-1}\fr{(1-q_0)}{q_0} -\fr{\sqrt{2q_0-1}}{q_0}].
\ee
Defining the red-shift z as ( $ z+1= \fr{R_0}{R}$), eq.(2.30) can be written as
\be
t=H_0^{-1}\int_0^{(1+z)^{-1}} \fr{dx}{[1-\Omega_0+\Omega_0x^{-1}]^{1/2}}
\ee
From the above equation we see that the age of the Universe is a decreasing
function of $\Omega_0$ : large $\Omega_0$ implies faster deceleration, which in turns
corresponds to a more rapidly expanding Universe in the past.\\
The age of the Universe provides a very powerful constraint to the value of
$\Omega_0$, and to the present energy density of the Universe. Independent
measurements suggest that
$$ t_0=10  \ \ \ to \ \ \  20 \ \ Gyr $$
In terms of $z$ and $\Omega_0$ one can write the age of the Universe as;\\
$$ t =\fr{2}{3}H_0^{-1}(1+z)^{-3/2},$$
for $\Omega_0=1 $
and
$$ t=H_0^{-1}\fr{\Omega_0}{2(1-\Omega_0)^{3/2}}[-\cosh(\fr{\Omega_0z-\Omega_0+2)}{\Omega_0(1+z)}+\fr{2(1-\Omega_0)^{1/2}}{(\Omega_0z+1)^{1/2}\Omega_0(1+z)}].$$
for $\Omega_0< 1$.
A closed universe has a maximum radius of $R_{max}=\fr{2q_0}{H_0(2q_0-1)^{3/2}}$  at
$ t_{max}=\fr{\pi}{2H_0\sqrt{\Omega_0}}$\ .\\
3.{\it Open model}\\
This is the case $k$ = -1. eq.(2.18) becomes
$$ \dt R^2=(\fr{\bt}{R}+1), \ \ \ \bt=\fr{2q_0}{H_0(1-2q_0)^{3/2}}. $$
The age of the Universe is
$$ t_0=\fr{q_0}{H_0(1-2q_0)^{3/2}}{[\fr{\sqrt{1-2q_0}}{q_0}-\ln\fr{(1-q_0+\sqrt{1-2q_0})}{q_0}]}. $$
Like in the Einstein-de Sitter model, the Universe in this model continues to expand forever.
\se{Radiation dominated (RD) Universe }
The solution of eq.(2.24) gives ($p=\fr{1}{3}\rh $)
\be
\rh\sim R^{-4}
\ee
and from eq.(2.18),
\be
R\sim t^{-1/2}
\ee
so that
\be
\rh\sim t^{-2}
\ee
In this case the curvature term is negligible in comparison with the second
term in (2.18). Therefore, $k$ = 0, 1, or -1 does not change the physical results.

The radiation  in this epoch is modeled by that of a black body.
The density of the black body radiation is related to its temperature by
\be
\rh_r=a T^4_r
\ee
where
$a=8.418\times 10^{-36}\rm gcm^{-3}\rm K^{-4}$.

According to eq.(2.33), the temperature of radiation is inversely proportional
to the scale factor of the Universe: (This also follows from the fact that the
black body radiation retains its spectrum during the expansion of the Universe)
\be
T_r\propto R^{-1}.
\ee
When $R$ is very small, that is, in the early Universe, $T_r$ can be very high.
Equations (2.28) and (2.33) give
\be
\fr{\rh_r}{\rh_m} = (\fr{\rh_r}{\rh_m})_0(\fr{R_0}{R}).
\ee
This formula shows that the ratio of the radiation and matter densities
is not invariant; rather, it decreases as the Universe expands.
Although the value $(\rh_r/\rh_m)_0$ is very small, being only $10^{-3}$; in
 the early Universe, i.e. that is when
$$\fr{R_0}{R} \gg 1\ , $$
we had
$$ \fr{\rh_r}{\rh_m} \gg 1. $$
Namely, the radiation was the dominant component of the Universe and its
temperature was
$$T_r=(T_r)_0(R_0/R) =T_{r0}(1+z)\ .$$
For $T_{r0}=2.7\rm K$ and $z\sim 10^3$, this corresponds to
$$ T_r > 2.7\times 10^3 \sim 3000 \rm K .$$
This phase of the Universe is called the radiation dominated (RD) Universe.

Since the Friedmann models are frequently used to interpret cosmological
observations, we will now derive some of the observable quantities in
these models:
\section{The cosmological tests}
\subsection{The red-shift}
Consider a galaxy $G_1$ at ($r, \theta, \phi$) emitting light waves towards us.
 Let us denote by $t_0$ the present epoch of observation. At what time
 should a light wave leave $G_1$ in order to arrive at $r=0$ at time $t=t_0$ ?.
 Since a light signal moves in a null geodesic, $ ds=0 $. Then the RW line
 element (eq.2.6)  gives us
 \be
 t =\pm\int \fr{R\ dr}{\sqrt{1-kr^2}}\ \ .
 \ee
 Since $r$ decreases as $t$ increases along the null geodesic we should
 take the minus sign in the above equation.\\
 Suppose that the light left $G_1$ at time $t_1$, hence from the above equation
 \be
 \int_{t_1}^{t_0}\fr{dt}{R(t)}=\int_0^{r_1}\fr{dr}{\sqrt{1-kr^2}}\ \ .
 \ee
Thus if we know $R(t)$ and $k$ we know the answer to our question.

Assume the wave crest was emitted
at $t_1$ and $t_1+\Delta t_1$ and received at $t_0$ and $t_0+\Delta t_0$,
respectively. Then
\be
 \int_{t_1+\Delta t_1}^{t_0+\Delta t_0 }\fr{dt}{R(t)}=\int_0^{r_1}\fr{dr}{\sqrt{1-kr^2}}\ \ .
 \ee
 If $R$ is a slowly varying function of time, i.e. it effectively remains
 unchanged over the small intervals $\Delta t_1$ and $\Delta t_0$
we get, by subtracting eqs.(2.40) and (2.41)
 \be
 \fr{\Delta t_0}{R(t_0)}=\fr{\Delta t_1}{R(t_1)}\ \ .
 \ee
i.e.
 \be
 \fr{\Delta t_0}{\Delta t_1}=\fr{R(t_0)}{R(t_1)}= 1 + z\ \ .
 \ee
The quantity $z$ defines a red-shift ($\lambda_1=c \delta t_1 , \lambda_0=c \delta t_0)$.
The wave length of the light wave increases by a factor $z$ in transmission
from $G_1$ to us, provided $ R(t_0) > R(t_1)$. Thus Hubble's
observations of the red-shift is explained if $R(t)$ is an increasing function of time.
This red-shift arises due to the passage of light through non-Euclidean
space time. It does not arises from the Doppler effect, since in our coordinate
 frame all galaxies have constant ($r, \theta, \phi$ ) coordinates. In non-Euclidean
space-time it is not possible to attach an unambiguous meaning to the relative
velocity of two objects separated by a great distance. Equation (2.43)
may be compared with the gravitational red-shift which is characterized by
the fact that if light from object B to A is red-shifted, the light from A
to B is blue shifted. In the present case, if light from A to B is
  red-shifted, that from B to A will also be red-shifted provided $R(t)$
  is an increasing function of time. We will therefore refer to the present
  red-shift as {\it cosmological red-shift}.
\subsection{ Luminosity distance ($D_L$)}
This is defined by \ \ [2]
\be
D_L=r_1R_0(1+z)\ \ .
\ee
{\it 1. Einstein-de Sitter model}\\
eqs.(2.39) and (2.27) give
\be
r_1 =\int_{t_1}^{t_0}\fr{dt}{R(t)}= \fr{1}{R_0}\int_{t_1}^{t_0}t_0^{2/2}t^{2/3}\ dt
\ee
\be
r_1 = \fr{3t_0}{R_0}[1-(t/t_0)^\fr{1}{3}]
\ee
Upon using eq.(2.29), this becomes
$$ r_1=\fr{2}{R_0H_0}[1-(1+z)^{-1/2}]$$
The luminosity distance becomes
$$D_L=\fr{2}{H_0}[(1+z)-(1+z)^{1/2}] $$
{\it 2. Closed Model}
\be
D_L= \fr{1}{H_0q_0^2}[q_0z+(q_0-1)(\sqrt{1+2zq_0}-1)]
\ee
{\it 3. Open Model}
\be
D_L= \fr{1}{H_0q_0^2}[q_0z+(q_0-1)(\sqrt{1+2zq_0}-1)]
\ee
Note that the equations for $D_L$ for the closed and open models are the same.
\subsection{The angular size}
We will study how the apparent angular size varies with red-shift in different
Friedmann models. We will assume that sources of a fixed size $d$ are observed
at different red-shifts. Thus a source at ($r, \theta, \phi$) with red-shift
$z$ will subtend at the observer at $r = 0$, the angle
\be
\Delta\theta_1=\fr{d}{r_1R(t_1)}=\fr{d(1+z)^{2}}{D_L}
\ee
$\Delta\theta_1$ is defined in terms of $z$ and $q_0$. It is interesting
to note that $\Delta\theta_1$ does not steadily decrease as $z$ increases.
 For $q_0=1/2 $
\be
\Delta\theta_1=\fr{d H_0(1+z)^{3/2}}{(1+z)^{1/2}-1}
\ee
A minimum value for $\Delta\theta_{1}$ occurs at $ z = 1.25 $:
$$\Delta\theta_{1\ min}= 6.75 H_0 d .$$
The cases $(q_0> 1/2$) are more involved [2].
\subsection{Source counts}
The number of astronomical sources with red-shifts between $z$ and $z+dz$
 is given by (applicable to all Friedmann models) [2]
\be
dN=\fr{4\pi r^2n(t)\ dr}{\sqrt{1-kr^2}}
\ee
\be
dN=\fr{4\pi}{H_0^3}[\fr{q_0 z+(q_0-1)(\sqrt{1+2q_0 z}-1)]^2}{q_0^4(1+z)^6\sqrt{1+2q_0z}} \bar{n} dz
\ee
where $\bar{n}=\fr{n}{R^3}$.
\subsection{Particle horizon}
The proper distance to the horizon in a RW space-time is given by
\be
d_H(t)=R(t)\int_0^t\fr{dt'}{R(t')}.
\ee
If $R(t)\propto t^n$, then for $n <1$, $d_H(t) $ is finite and is equal
to $\fr{t}{(1-n)}$\ , i.e. in spite the fact that all physical distances approach
zero as $R\rightarrow 0$,
the expansion of the Universe precludes all but a tiny fraction of the
volume of the  Universe
from being in casual contact. This is a vexing feature of the Standard
Model.
A more general expression for the $d_H$ [5], is
\be
d_H(t)=\fr{1}{H_0(1+z)}\int_0^{(1+z)^{-1}}\fr{dx}{[x^2(1-\Omega_0)+\Omega_0x^{(1-3w)}]^{1/2}}
\ee
From this expression we see that if $w <-1/3 $, the integral
will diverge and the horizon distance will be infinite.
The present horizon distance is given by
$$ d_H=\fr{1}{H_0}
\left \{
 \begin{array}{c} 2\ \ \ \ \  \ \  \ \ \ k=0,\ \ \ \ \  q_0=1/2   \\
\fr{2}{\sqrt{2q_0-1}}\sin^{-1}\sqrt{\fr{2q_0-1}{2q_0}}  \ \ \ \ \ \  \ k=1, \ q_0 >1/2\\
\fr{2}{\sqrt{1-2q_0}}\sinh^{-1}\sqrt{\fr{1-2q_0}{2q_0}}\ \ \ \ \  k=-1, q_0 < 1/2\\
 \end{array}
\right\} $$
 The existence of a finite $d_H$ means that the Universe has a particle horizon.
 \subsection{Event horizon}
 A related question to that posed at the beginning of subsection 2.10.1 is whether
a light signal sent out at the present
 time ($t_0$) reaches all points of the Universe before its end at time $t_E$.
 Since  light travels a maximum coordinate distance
 $$d_{EH} =\int_{t_0}^{t_E}\fr{dt}{R(t)} $$
 there exists an {\it event horizon} if this is smaller than $\pi$ or $\infty$:
 we shall never learn any thing about events which at the present time are situated
 at a distance greater than the above distance [5]
\se{References}
$[1]$ Creation of the Universe, Li. F. Zhi , 1989 (World Scientific)\\
$[2]$ Introduction to Cosmology, J. V. Narlikar, 1983 (Jones \& Bartlett)\\
$[3]$ Introduction to Cosmology, M. Roos, 1994 (John Wiley)\\
$[4]$ The Isotropic Universe, D. J. Raine, 1981 (Adam Hilger)\\
$[5]$ M.S.Turner and Kolb, The Early Universe (1990)(Addison-Wesily Publishing Company)\\
$[6]$ A. H. Guth, Phy. Rev.D23, 347(1981)
\chapter{Thermal History of the Universe}
\se{Thermodynamics of the Universe}
\subsection{Introduction}
The first theoretical basis of the evolutionary view is thermodynamics.
Applying the requirement of thermal equilibrium to the Universe, we have
to say that the general tendency of the cosmic evolution is for the
temperature to be the same all over.
It was pointed out by Helmholtz in 1854 that the whole Universe will eventually be
in a state of uniform temperature and will be falling in the state of thermal
death (eternal rest).
According to the thermodynamics any temperature difference between two systems
will approach zero. Using thermodynamics, it was proved that for an expanding Universe,
even if the initial temperature is the same, a temperature difference may
still be generated. Consider a spherical region R  and assume that the
matter in R  has already reached  equilibrium at the beginning. Roughly speaking, there
are two kinds of matter in the Universe, one is baryonic and the other is radiation.
Let $\rh_m$ and $\rh_r$ be the mass densities of matter and radiation respectively.
Their equations of state are [1]
\be
p_r=\fr{1}{3}\rh_r
\ee
\be
\rh_m\simeq\ m\ n+\fr{3}{2}nT
\ee
where $n$ is the number density of the nonrelativistics particles and $m$ their rest mass.
Here we have assumed that each particle has three degrees of freedom. The expansion
of the Universe should be adiabatic. No exchange with the exterior for the
exterior does not exist in the Universe. The expansion of the region R
which is  typical of
the Universe, must also be adiabatic. Even though an exterior region for R exists,
there is no difference between R and its exterior because the Universe
is uniform throughout. The adiabatic expansion of a system satisfies
\be
dE=-pdV\ ,
\ee
where $E$ is the total energy of the system.
This formula constitutes the basis for the thermodynamics of an expanding
Universe. We now proceed to investigate the consequences of these equations.
\subsection{Radiation under adiabatic expansion}
The total energy density in the region R is
\be
E_r=V\rh_r, \ \ \ V\sim R^3.
\ee
Substituting eq.(3.4) in (3.3) gives
\be
d(R^3 \rh_r) =-p_rdR^3.
\ee
Then using eq.(3.1), this becomes
\be
d(R^3\rh_r) =-\fr{1}{3}\rh_rdR^3
\ee
or
\be
R^3d\rh_r+\rh_rdR^3=-\fr{1}{3}\rh_rdR^3
\ee
i.e.
\be
\fr{d\rh_r}{\rh_r}=-\fr{4}{3}\fr{dR^3}{R^3}.
\ee
The solution of this equation is
\be
\rh_r\propto R^{-4}.
\ee
According to the thermodynamics of radiation the relation between $\rh_r$ and
 the temperature of radiation $T_r$ is
\be
\rh_r\propto T^4 .
\ee
From eqs.(3.9) and (3.10) we immediately get
$$T_r\propto R^{-1}\ . $$
This result shows that as the Universe expands the radiation
temperature falls in inverse ratio to the scale factor
$R$\ , provided only radiation exists.
\subsection{Matter under adiabatic expansion}
Assuming the presence of matter only we apply eq.(3.3) to obtain
\be
d(R^3\rh_m) = -p_mdR^3,
\ee
where $p_m$ is the pressure of matter,
\be
p_m=n T_m.
\ee
Substituting eqs.(3.12) and (3.2) in (3.11), we then find
\be
d(R^3mn)+d(R^3\fr{3}{2} nT_m)=-nT_mdR^3,
\ee
The total number of particles $N=nV$ is conserved within $V$,
and so $n\sim R^{-3}$.
Thus, eq.(3.13) now becomes
\be
\fr{3}{2}d(R^3nT_m)=-nT_mdR^3\ .
\ee
Using $N=nV$, this becomes
\be
\fr{3}{2}\fr{dT_m}{T_m} = -\fr{dR^3}{R^3}\ ,
\ee
with the solution
\be
T_m \propto R^{-2}\ .
\ee
This  result shows that, as the Universe expands, the matter temperature
$T_m$ also decreases but in a different manner to radiation.
\subsection{Generation of temperature differences}
$T_r$ and $T_m$ vary with $R$ in different ways. Hence  during the process of cosmic
expansion it is  impossible for $T_r$ and $T_m$ to be always equal.
Even if initially $T_r$ and $T_m$ were equal,
later on, with expansion, we will have $ T_r>T_m $.
Therefore, the cosmic expansion  saves the Universe from  the final outcome
of thermal death.
According to thermodynamics, a system with thermal equilibrium has the same temperature
for all  the various components, but if radiation and matter always keep the same
temperature, how can temperature difference ever appear ?\\
In a system with complete  thermal equilibrium all components must have the same temperature.
However, a certain time is required to reach equilibrium, interaction between radiation
and matter must be carried on for a length of time before the two can achieve the same
temperature. If the time required to achieve  uniform temperature is longer  than the
time scale of the cosmic expansion, then there will never be thermal equilibrium
between radiation and matter. In this case, it is reasonable to separately solve the
thermodynamics  equations for the two components. Due to cosmic expansion, matter is not
in a state  of complete thermal equilibrium. Radiation and matter
are separated from thermal equilibrium because there has not been sufficient time for
the two components to achieve mutual equilibrium.
\subsection{Equilibrium  thermodynamics}
Today the radiation in the Universe is comprised of the $2.75 \rm K$
microwave photons and the cosmic sea of $1.96 \rm K$ relic neutrinos. Because the early Universe
was to a good approximation in thermal equilibrium, there should have been  other relativistic
particles present, with comparable abundance.
The number density $n$, energy density $\rh$, and pressure of a dilute, weakly interacting
gas of  particles with $g$ internal degrees of freedom, in thermal equilibrium
at temperature $T$, are given by [5]
\be
n=\fr{g}{(2\pi)^3}\int f(q)\ d^3q,
\ee
\be
\rh=\fr{g}{(2\pi)^3}\int E(q)f(q)\ d^3q,
\ee
and
\be
p=\fr{g}{(2\pi)}\int\fr{q^2}{3E}f(q)d^3q
\ee
where $E^2=q^2+m^2$ and $f$ is the Fermi-Dirac (FD) or Bose-Einstein (BE)
distribution for species in kinetic equilibrium.
\be
f(q)=(\exp\fr{E-\mu}{T}\pm 1)^{-1},
\ee
where $\mu$ is the chemical potential of the species, +1 is chosen for FD
species and $-1$ for BE species.
It follows that
\be
\rh=\fr{g}{2\pi^2}\int_m^{\infty}\fr{(E^2-m^2)^{1/2}E^2\ dE}{\exp(\fr{E-\mu}{T}) \pm 1 },
\ee
\be
n=\fr{g}{2\pi^2}\int_m^{\infty}\fr{(E^2-m^2)^{1/2}E\ dE}{\exp(\fr{E-\mu}{T})\pm 1},
\ee
\be
p=\fr{g}{6\pi^2}\int_m^{\infty}\fr{(E^2-m^2)^{3/2}\ dE}{\exp\fr{E-\mu}{T}\pm 1},
\ee
For $T\gg m$ (relativistic limit) and  $T\gg \mu$
\be
\rh=\left\{
\begin{array}{c}\fr{\pi^2}{30}g T^4,\  \ \ BE\\
\fr{7}{8}\fr{\pi^2}{30}g T^4\ \ \ FD
\end{array}
\right\}
\ee
and
\be
n=\left\{
\begin{array}{c}\fr{\xi(3)}{\pi^2}g T^3\ \  \ BE\\
\fr{3}{4}\xi(3)g T^3  \ \ \ FD
\end{array}
\right\} \ee $\xi(3)=1.20206 $ is the Riemann zeta function. In
the nonrelativistics limit
\be
n=g(\fr{mT}{2\pi})^{3/2}\exp-\fr{(m-\mu)}{T}\ ,
\ee
\be
\rh = mn,
\ee
\be
p=nT\ll \rh\ ,
\ee
for BE and FD species.
For  relativistic particles, eqs.(3.21) and (3.23) give
\be
\rh_r=\fr{\pi^2}{30}g_*T^4
\ee
\be
p_r=\fr{1}{3}\rh_r=\fr{\pi^2}{90}g_* T^4
\ee
where  $g_*$, the total number of effectively massless degrees of freedom
($m_i\ll T$), is given by
\be
g_*=\sum_{i=boson}g_i(\fr{T_i}{T})^4+\fr{7}{8}\sum_{i=fermion}g_i(\fr{T_i}{T})^4
\ee
and $T_i$ is the temperature of species $i$ that may not have the same temperature as
photons.
The factor $\frac{7}{8}$ accounts for the difference in FD and BE statistics.
\subsection{Entropy}
Through most of the history of the Universe (in particular the early Universe)
the reaction rates of particles  in the thermal bath, $\Gam_{int}$, were much
greater than the expansion rate, $H$, and local thermal equilibrium should have
been maintained. In this case the entropy per comoving volume element remains constant.
The entropy in a comoving volume provides a very useful fiducial quantity
during the expansion of  the Universe. The second law of thermodynamics implies
\be
TdS\equiv d(\rh V)+pdV=d[(\rh+p)V]-Vdp
\ee
The integrability condition [5]
\be
\fr{\partial^2 S}{\partial T\partial V}=\fr{\partial^2 S}{\partial V\partial T}
\ee
relates the energy density and the pressure
\be
T\fr{dp}{dT}=\rh+p
\ee
or
\be
dp=\fr{\rh+p}{T}dT\ .
\ee
Equation (3.32) becomes
\be
dS=\fr{1}{T}d[(\rh+p)V]-\fr{(\rh+p)V}{T^2}dT
\ee
\be
dS=d[\fr{(\rh+p)V}{T}+const.].
\ee
The entropy per comoving volume is (up to an additive constant)
\be
S=\fr{R^3(\rh+p)}{T}
\ee
The general relativistics field equations require
\be
dS=0
\ee
 This results implies that  in thermal equilibrium the entropy, per comoving volume,
 is constant.
 It is useful to define the entropy density  $s$
 \be
 s\equiv\fr{S}{V}=\fr{\rh+p}{T}.
 \ee
At high temperature the entropy density is dominated by the contribution of the relativistic
 particles, so that to a very good approximation\footnote{$s$ is also proportional to
 the density of relativistic particles and in particular\\ $s=1.8g_{*s}n_\gam$, $n_\gam=$ photon density}
 \be
 s=\fr{2\pi^2}{45}g_{*s} T^3
 \ee
 \be
 g_{*s}=\sum_{i=boson}g_i(\fr{T_i}{T})^3+\fr{7}{8}\sum_{i=fermion}g_i(\fr{T_i}{T})^3.
 \ee
 During thermal equilibrium all particle species had a common
 temperature.
 Whenever a particle species becomes non relativistic and decouples its entropy is
 transferred to the other relativistic particle species still present in the thermal
 plasma, causing $T$ to decrease slightly  less slowly.
 Massless particles that decoupled from the heat bath will not share in the entropy
transfer as the temperature drops below the mass threshold of a species. Instead,
the temperature of the  decoupled massless particles  scales as $R^{-1}$\ .
\se{The early radiation era}
The Friedmann  equation (2.18) for the radiation era is
\be
\fr{\dt R^2}{R^2}=\fr{8\pi G}{3}\rh.
\ee
During this epoch $ \rh \propto R^{-4} $, so that
\be
R\sim t^{1/2}.
\ee
The energy density of the dominant black-body radiation in this era obeys the equation
\be
\rh =\fr{g}{2} a T^4,\ \ \ \ a=\fr{\pi^2}{15}\ ,
\ee
where $g$=2 for photons. Therefore, it follows that
\be
\fr{\dt R}{R}=-\fr{\dt T}{T}
\ee
and that
\be
(\fr{\dt T}{T})^2=\fr{8\pi G}{3}\fr{g_*\pi^2}{30} T^4
\ee
The solution of this latter equation  is
\be
T=(\fr{3}{16\pi G g_*a})^{1/4} t^{-1/2}= (\fr{45}{16\pi^3 Gg_*})^{1/4} t^{-1/2}\ ,
\ee
giving
\be
T(K)= \fr{1.805}{g_*^{1/4}} t^{-1/2} (sec)
\ee
and
\be
\rh = \fr{3}{32\pi G}t^{-2}\ .
\ee
To determine which types of particles will be in thermal equilibrium at a given
temperature one proceeds as follows: we note that if the particle's number density is $n$,
its velocity is $v$ and its reaction cross section is $\sigma$, then
the average reaction rate is given by [6]
\be
\Gam= <n\sigma v> \ee averaged over the thermal distribution. Some
relevant reaction are: $e^++e^-\rightarrow \mu+\bar{\mu},
e^++e^-\rightarrow \nu+\bar{\nu}$, $q+\bar{q}\rightarrow
q+\bar{q}$, $\ell+\bar{\ell}\rightarrow q+\bar{q}$ where $q, \ell$
stand for quarks and leptons respectively.
 A given species of particles will remain in  thermal equilibrium so long as
this reaction rate is sufficiently high compared to the age of the Universe, i.e.
if
\be
\Gam \gg H=\fr{1}{2t}
\ee
Hence initially when $n$ and $v$ are very large we expect all particles
to be in equilibrium, but once the temperature drops, and the energy of
a particular particle falls below the relevant reaction threshold, $\sigma$
will vanish and the particle will cease to be in equilibrium. Therefore,
when  $\Gam > H \ ( \Gam < H ) $ the particle species couple (decouple)
from the thermal bath.\\
For reactions mediated by massless gauge bosons\footnote{ e.g.$ \ e^-+e^+\rightarrow
\gam+\gam\ \ , e^-+e^+\rightarrow \mu^-+\mu^+$}[5] $\Gam\sim n\sigma|v|\sim \alf^2T$,
during the RD epoch $ H\sim \fr{T^2}{m_{pl}}$, so that
$\fr{\Gam}{H}\sim\fr{\alpha^2m_{\rm Pl}}{T}$,
($\sqrt{4\pi\alpha}=g=$the gauge coupling strength).\\
Therefore, for $ T < \alf^2 m_{pl}\sim 10^{16} \rm GeV $, the reactions
are occurring rapidly, while for $ T> \alpha^2m_{pl}\sim 10^{16} \rm GeV $, they
are effectively ``frozen out." \\
For reactions mediated by massive gauge bosons\footnote{e.g. $e^-+p\rightarrow n+\nu,
e^-+e^+\rightarrow\mu^-+\mu^+\ \ , \ell+\bar\ell\rightarrow q+\bar q$}: $\Gam\sim n\sigma|v|\sim G^2_X T^5 $ and
$ \fr{\Gam}{H}\sim G^2_X m_{pl}T^3$, where $G_X\sim\fr{\alpha}{m_X^2}$ and
$m_X$ is the mass of the gauge bosons.
Thus for $ m_X > T > G^{-2/3}m_{pl}^{-1/3}\sim (m_X/100\rm GeV)^{4/3}MeV$,
the reactions are occurring rapidly, while for $ T<(m_X/100\rm GeV)^{4/3}MeV$,
they are effectively frozen out.
\se{Neutrino decoupling}
For $T\gg 1\rm MeV$, neutrons would be in equilibrium with the rest of the
Universe via the weak interaction processes ($e^++e^-\rightarrow\nu+\bar\nu$\ ,
$\nu+e^-\rightarrow \nu+e^-\ \ , e^-+p\rightarrow\nu+n$\ , etc.).
The weak interaction cross section $\sigma$ is given by
\be
\sigma\simeq  G_F^2T^2
\ee
where $G_F$ is the Fermi's constant. The number density of neutrinos is $\propto T^3$.
The reaction rate for neutrinos ($\Gam_{wi}$) therefore falls with decreasing temperature as $T^5$.
Hence eq.(3.51) gives
\be
\fr{\Gam_{wi}}{H}\propto T^3
\ee
When eq.(3.52) is not fulfilled, neutrinos decouple (freeze out) from all
interactions and begin a free expansion. The decoupling of $\nu_\mu$ and $\nu_\tau$
occurs at $3.5 \rm MeV$ whereas the $\nu_e$ decouples at $2.3 \rm MeV$. At decoupling, the
neutrinos are still relativistic and thus their energy distribution is given by
the Fermi distribution and their average temperature equals that of photons.\\
 As the Universe cools and the energy approaches $0.8 \rm MeV$, the reactions converting
the protons into neutrons stop but the neutrons also decay into protons by
the beta decay, $n\rightarrow p+e+\bar{\nu}$,
and therefore the ratio of protons to neutrons increases.
The mean life of the neutron is 889s and in comparison with the age of the
Universe which at this time is a few tens of seconds the neutrons are
essentially stable.
The electromagnetic cross sections are $\propto T^{-2}$, and the reaction rate ($\Gam_{em}$)
is then $\propto T$ so that
 \be
 \fr{\Gam_{em}}{H}\propto \fr{1}{T}
 \ee
 Equation (3.52) is satisfied for all temperatures, so, in contrast to the
 weak interactions, the electromagnetic interactions never freeze out.
 The reaction
 \be
 e^- + e^+\rightarrow \gam+\gam
 \ee
 creates new photons with energy $0.51 \rm MeV$. This is higher than the ambient
 photon temperature at that time, so the photons population gets reheated.
 To see this consider eq.(3.41)
 \be
 s=\fr{2\pi^2}{45}g_{*s} T^3,
 \ee
 Since entropy is conserved throughout, this is only possible if $g_{*s}T^3$ remains
 constant. By applying this argument to the situation when positrons and most
 of the electrons disappear by annihilation below 0.2 MeV. We denote quantities
 above this energy by + , and below it by -- . Above this energy the particles
 in thermal equilibrium are $\gam, e^-, e^+ $. Then the entropy is
 \be
 s=\fr{11}{3}aT^3_+\ .
 \ee
 Below this energy only photons contribute the factor $g{_*s}=2$\ .
 Hence,
 \be
 T_-=(\fr{11}{4})^{1/3}T_+=1.40 T_+
 \ee
 The number density of neutrinos is related to that of photons as
 \be
 n_\nu=\fr{3}{4}.\fr{4}{11}n_\gam
 \ee
 When electrons become slow enough, they are captured into atomic orbits by
 protons, forming  stable hydrogen atoms ($B.E= 13.6 \rm eV$). Actually  the formation of hydrogen atom
 occurs at $0.3 \rm eV$, because the released binding energy reheats the remaining
 electrons, and also because the large amount of entropy in the Universe favors
 free protons and electrons. When this recombination is completed, the photons
 find no more free electrons to scatter against, thus the photons decouple and the
 Universe becomes transparent to radiation [3].
\se{ Nucleosynthesis}
We left the story of the decoupling nucleons at the time when the weak interaction
ceased and the conversion of protons to neutrons stopped because the energy in
the thermal bath dropped below $0.8\rm MeV$. The neutrons and protons were then non-relativistic,
so their number densities were each given by the Maxwell-Boltzmann distributions.
For nuclear species $A(Z)$ with mass number $A$ this is given by
 \be
 n_A=g_A(\fr{m_AT}{2\pi})^{3/2}\exp(\fr{\mu_A-m_A}{T})
 \ee
 where $\mu$ is the chemical potential of the species. At equilibrium this is related
 to the chemical potential  of the proton ($\mu_p$) and neutron ($\mu_n$) by,
 \be
 \mu_A=Z\mu_p+(A-Z)\mu_n
 \ee
 By eliminating the $\exp(\mu_A/T)$ term, one can write [5]
\be
n_A=g_A A^{3/2}2^{-A}(\fr{2\pi}{m_NT})^{3(A-1)/2}n_p^Zn_n^{A-Z}\exp(B_A/T)\ \ .
\ee
\be
B_A\equiv Zm_p+(A-Z)m_n-m_A\ \ ,
\ee
 So the neutron to proton ratio is $$ n_n/n_p=\exp(-Q/T)\ , $$ where $Q=m_n-m_p$\ .
 At energies of the order of $(m_n-m_p)=1.293 \rm MeV$ or less, this ratio is dominated by the
exponential. Thus at $T = 0.8\rm MeV$ one finds that the ratio has dropped from 1 to 1/5.
Already at a few $\rm MeV$, nuclear fusion reactions start to build up light elements.
For a typical bound nucleus $A$ with atomic mass $A$ and number density
$ n_A$ the mass fraction of the various nuclear species $A(Z)$ is given by
\be
X_A\equiv \fr{n_AA}{n_N}
\ee
where $n_N$ is the total number density of nucleons.
From eq.(3.63) it follows that

$$X_A=g_A[(\zeta(3)^{A-1}\pi^{(1-A)/2 }2^{(3A-5)/2}]A^{5/2}(T/m_N)^{3(A-1)/2}$$
\be
\times\eta^{A-1}X_p^ZX_n^{A-Z}\exp(B_A/T)
\ee
where $\eta\equiv\fr{n_N}{n_\gam}=2.68\times 10^{-8}(\Omega_Bh^2)$\ ,
$\Omega_B$ being the baryonic fraction of the critical density.
 The equilibrium abundance of deuterons is given by :
 \be
X_d= 16.3(\fr{T}{m_N})^{3/2}\eta X_p X_n\exp( \fr{B_d}{T})\ .
 \ee
  Since $ B_d$, the deuteron binding energy ($=2.22 \rm MeV)$, is low,
$ X_d$ is not high enough to start fusion reactions leading to $ ^2H, ^3H $
 and $^4He$  unless $T$ drops to less than $ 10^{9} \rm K$.
 Although the deuterons are formed in very small
quantities, they are of crucial importance to the final composition of matter.
Photons of energy $2.22 \rm MeV$ or more photodisintegrate the deuterons into  free protons
and neutrons.

Consider the above process at equilibrium, and define the relative
abundance by
\be
\fr{X_nX_p}{X_d}=\fr{4}{3}\fr{(2\pi T)^{3/2}}{n_B(2\pi)^3}(\fr{m_N^2}{m_d})^{3/2}\exp(\fr{B_d}{T}).
\ee
where $m_N=m_B$ is the nucleon mass.

The above equation tells us that as the Universe cools the
equilibrium shifts in favor of $d$ over $p$ and $n$
 at $T\simeq 10^9 \rm K$.
 We denote the temperature $T_N$, when the above ratio equals
to unity, as the nucleosynthesis temperature.

All evidence suggest that the number density of baryons (nucleons) is
today very small. In particular, we are able to calculate it up to
a multiplicative factor $\Omega_B h^2$, [3]
\be
n_B=\fr{\rh_B}{m_B}=\fr{\Omega_B\rh_c}{m_B}\simeq 1.13\times 10^{-5}\ \Omega_Bh^2\rm cm^{-3}.
\ee
The parameter $\Omega_B$ cannot be very much larger than unity, because that
would close the Universe too  fast.

Detailed calculations show that deuterons production becomes
thermodynamically favorable only at energy of magnitude $0.07 \rm
MeV$. Other nuclear reactions also commence at a few $\rm MeV$.
The $ ^3He $ is produced in the reactions
 $$ d+d\rightarrow ^3He+n,$$
$$ p+d\rightarrow ^3He+\gam,$$ with the binding energy $7.72 \rm
MeV$. This reaction is hampered by the large entropy per nucleon,
so it becomes thermodynamically favorable only at $0.11 \rm MeV$.
The $^3H$ is produced in the fusion reactions $$ n+d\rightarrow
^3H+\gam,$$
 $$ d+d\rightarrow ^3H+p,$$
 $$ n+^3He\rightarrow ^3H+p,$$
with the binding energy $8.48 \rm MeV$. A very stable nucleus is
the $^4He$ with a very large binding energy of $28.3 \rm MeV$.
Once its production is favored by the entropy law, at about $0.28
\rm MeV$, there are no more $\gam$-rays sufficiently energetic to
photodisintegrate it. $^4He$ is mostly produced in the reaction
\be
d+d\rightarrow ^4He+\gam
\ee
However $^3He$ and $^3H$ production is preferred over deuteron fusion, so $^4He$
is only produced in a second step when these nuclei become abundant.
The reactions are then
$$ n+^3He\rightarrow ^4He +\gam,$$
$$ d+^3He\rightarrow ^4He +p,$$
 $$ p+^3H\rightarrow ^4He+\gam,$$
$$ d+^3H\rightarrow ^4He+n  . $$
 At energy of magnitude $0.1 \rm MeV$  when the temperature is $1.2\times 10^9 \rm K$ and the time elapsed since the
big bang is a little  over 2 minutes, the beta decay of neutrons already noticeably
converts neutrons to protons. At this point the $ n_n/n_p$ ratio has reached its final
value.
\be
\fr{n_n}{n_p}\simeq \fr{1}{7}.
\ee
 These remaining neutrons have no time to decay before they fuse into deuterons
and subsequently into $^4He$. There they stay until today because bound neutrons
don't decay. The same number of protons as neutrons go into $^4He$, and the remaining
free protons are the nuclei of future hydrogen atoms. Thus the end result of nucleosynthesis
taking place between 100 and 700 seconds after the big bang is a Universe  composed
almost entirely of hydrogen and  helium nuclei. \\
Why not heavier nuclei ? It is an odd circumstance of nature that although there exist
stable nuclei
composed of 1, 2, 3 and 4 nucleons, no nucleus of  $A=5$ exits, and no stable
one with $A=8$. In between these gaps, there exist the unstable nuclei $^6Li, ^7Be$
and the stable $^7Li$. Because of these gaps and because $^4He$ is so strongly
bound, nucleosynthesis essentially stops after $^4He$ production. Only small minute
quantities of stable nuclei $^2He$, $^3He$ and $^7Li$ can be produced. The relic
abundance of the light elements bears quite an important testimony of the Big Bang.
The number of $^4He$ is clearly  half the number of protons as neutrons
go into $^4He$. Thus the excess number of protons going into the formation
of  hydrogen is $n_p-n_n$.
Usually one quotes the ratio of mass in $^4He$ to total mass in $^1H$ and $^4He$, [3]
\be
Y_4\equiv\fr{^4He}{^4He+^1H}\simeq 0.25.
\ee
 At present, the best estimate of $Y_4$ from observational data is in the range
0.22 -- 0.24 with no directional variation. This is a strong support of the big bang
hypothesis.
The helium mass abundance $Y_4$ depends sensitively on several parameters.\\
(i) The quantity $\eta$ in eq.(3.69) $\propto \Omega_B$. If the $n_B$ increases
$\Omega_B$ and $\eta$ also increase, and the entropy per baryon decreases. Remembering
that the large entropy per baryons was the main obstacle to early deuteron
and helium production, the consequence is that helium production can start earlier.
But then the neutrons would have had less time to beta decay, so the $n_n/n_p$ ratio
would be larger than $\frac{1}{7}$. It follows that more helium will be produced: $Y_4$ increases.\\
(ii) An increase in the neutron mean life implies a decrease in the reaction rate
$\Gam_{wi}$\ .
Hence a longer mean life implies a higher decoupling temperature and an earlier
decoupling time. As we have already seen, an earlier start of helium production leads
to an increase in $Y_4$.\\
(iii) The expansion rate $H \propto \sqrt{g_*}$ which in turn depends on
the number of neutrino families $N_\nu$. If there were more than 3 neutrino families
$H$ would increase .
Similarly if the number of neutrinos were very different from the number
of antineutrinos, contrary to the assumption in standard model, $H$ would also
increase.
From the nucleosynthesis  value of $\eta$ one can obtain an estimate of the baryon
density parameter $\Omega_B$. Combining eqs.(3.69) and  (3.70) one gets
\be
\Omega_B=\fr{n_B m_B}{\rh_c}=\fr{n_\gam m_B}{\rh_c}\eta\simeq 3.65\times 10^7\eta h^{-2},
\ee
The observational data [3]  constrains $\eta$ to
\be
2.8<10^{10}\eta<4.2 .
\ee
 This limit on $\eta $ gives:
 $$ 0.010  <  \Omega_B h^2 < 0.015 .$$
 Note that $h$ ranges from 0.5 to 0.85 [3] which implies
 $$ 0.01 < \Omega_B < 0.05 .$$

 Thus we arrive at the very important conclusion that there is too little baryonic
 matter to close the universe. Either the Universe is then indeed open, or there
 must exist other non-baryonic matter.
 \se{The decoupling of photons}
The relative abundance of free electrons of number density $n_e$, free
protons of number density $n_p$ and
free neutral $H$-atoms of number density $n_H(=n_B-n_e$) in thermal equilibrium
at a given temperature is determined by
 \be
 \fr{n_en_p}{n_H} = (\fr{m_e T}{2\pi})^{3/2}\exp (-\fr{B_H}{T})
 \ee
where $ B_H = 13.59 \ \rm eV $ is the hydrogen binding energy.
 The ionization fraction is given by
 $$x = \fr{n_e}{n_B}  \ \  ,  $$
so that
 $$ \fr {x^2}{1-x} = \fr{1}{n_B} (\fr{m_e T}{2\pi})^{3/2} \exp (-\fr{B_H}{T}) .$$
 For $ \Omega_0 h^2 = 0.1\ [2]  , x = 0.003 $ , at $T = 3000 \rm K $.
 Thus by this time most of the free electrons have been removed from the cosmological
 brew, and as a result the Universe becomes effectively transparent to radiation.
 This era signifies the beginning of a new phase when matter and radiation become
 decoupled. This phase continues to the present epoch.
 During this phase each photon frequency is red-shifted as $ R^{-1} $, the number
 density is $ \propto R^{-3} $ and the temperature is $\propto R^{-1}.$
 A background radiation of temperature $T\simeq 3 \rm K$ therefore means that the
red-shift at this epoch when matter decoupled from radiation was $z\approx 10^3 $.
 The opaqueness of the Universe prevent us from ``seeing" directly beyond the
 red-shift of $z\sim 10^3 $. Thus any evidence of the big bang  must come
 indirectly. The abundance of light nuclei and the MBR provide us with the
 only means of checking the very early history of the Universe.
\se{References}
$[1]$ Creation of the Universe, Li. F. Zhi, 1989 (World Scientific)\\
$[2]$ Introduction to Cosmology, J. V. Narlikar, 1983 (Jones \& Bartlett)\\
$[3]$ Introduction to Cosmology, M. Roos, 1994 (John Wiley)\\
$[4]$ The Isotropic Universe, D. J. Raine, 1981 (Adam Hilger)\\
$[5]$ E.W. Kolb and M. S. Turner, The Early Universe (1990) (Addison-Wesily Publishing Company)\\
$[6]$ P.D.Cllins,A.D.Martin, E.J.Squires, Particle Physics and Cosmology 1989,
Jon Wiley and Sons, Inc.
\chapter{Nonstandard Cosmological Models}
\se{Background}

From eqs.(2.18) and (2.20) we see that if we want a static solution of Einstein's equations, that
is, one in which $\dt R=0$, we must have $\rh+3p=0$, which is somewhat
unphysical, because, if $\rh > 0$ then $p < 0$, and if $p=0$ then
$\rh=0$. Therefore Einstein modified his equations by adding
the so-called {\it `cosmological term'} to the field equations, as follows
\be
{\cal{R}}_{\mu\nu}- \fr{1}{2} {\cal {R}}g_{\mu\nu}-\Lambda g_{\mu\nu}=8\pi G T_{\mu\nu}\ ,
\ee
where $\Lambda$ is the cosmological constant.
Hence  eqs.(2.14-16) in (4.1) give
\be
\fr{\dt R^2}{R^2}+\fr{k}{R^2}=\fr{8\pi G}{3}\rh+\fr{\Lambda}{3}\ ,
\ee
and
\be
\fr{\dd R}{R}=-\fr{4\pi G}{3}(\rh+3p) +\fr{\Lambda}{3}.
\ee
The energy conservation law (eq.(2.2)) takes the form
\be
\dt \rh+3\fr{\dt R}{R}(\rh+p)=-\fr{\dt\Lambda}{8\pi G}\ ,
\ee
so that the entropy $S$, defined in
\be
dE+\rh dV=dQ\equiv TdS,
\ee
is no longer constant. However, in the standard model the entropy is constant
and that was considered to be one of the problems of the standard model.
This is because the entropy of the Universe, today, is unexplainably
very huge.
Therefore, a variable cosmological term gives rise to an increasing entropy in
conformity with the second law of thermodynamics. This is true if the cosmological
constant decreases with expansion.

A more general relation between $k$ and $\Omega_0$ is given by
\be
\Omega_0(1+3w)=2q_0+\fr{2\Lambda}{3H_0^2}\ ,
\ee
and
\be
\fr{3}{2}\Omega(1+w)-q_0-1=\fr{k}{R_0^2H_0^2}\ ,
\ee
where $w$ is defined by the equation of the state  $p=w \rh$, and
the subscript 0 denotes the present day quantities.
Consider a Universe for which $k\ge 0$. It then follows for $w=0$ that
\be
\Omega_0\ge 1-\fr{\Lambda_0}{3H_0^2}.
\ee
Thus we see that the cosmological constant changes the simple relation
between $k$ and $\Omega$, see eq.(2.21). For instance, a flat universe is no longer
 characterized by $\Omega=1$.
Note that although $\Lambda$ is exceedingly small, the term $\fr{\Lambda}{3H_0^2}$ may be
of the order of unity. The value of $\Omega=1$ is preferred on theoretical
grounds, but the observational values are mostly much smaller than 1.

If we demand that $R(t)=R_0$=constant, and  $p=0$, we get
\be
\rh=\fr{\Lambda}{4\pi G}\ \ , \ \ \ k=\fr{\Lambda}{ R_0^2}\ .
\ee
Thus $\Lambda$ must be positive and therefore $k=+1$. This is the Einstein's static
model. Later on, Einstein regretted the addition of this term when he knew about
the expansion of the Universe.

Many dynamical solutions with the cosmological term were studied by Lemaitre
and they were accordingly known as {\it Lemaitre models.}
 In recent years other motivations have been found for introducing a cosmological
 term. Introducing this term is like introducing a fictious `fluid' with energy
 momentum tensor $T'_{\mu\nu}$ given by
 \be
 T'_{\mu\nu}=(\rh'+p')u_\mu u_\nu-p'g_{\mu\nu}=\fr{\Lambda}{8\pi G}g_{\mu\nu}\ ,
 \ee
 so that the energy density and pressure of this fluid are given by $\rh'=\fr{\Lambda}{8\pi G}, \ \ \ p'=-\fr{\Lambda}{8\pi G} $.
Hence eq.(4.1) can be written as
 \be
{\cal{R}}_{\mu\nu}- \fr{1}{2} {\cal {R}}g_{\mu\nu}=8\pi G(T_{\mu\nu}+T'_{\mu\nu}).
\ee
\se{Limits on the cosmological constant}
From eq.(4.4) one gets [32],
\be
q_0=\fr{\Omega_0}{2}-\Lambda/3H_0^2
\ee
$$ |q_0-\Omega_0/2|=|\Lambda/3H_0^2|$$
Though $\Omega_0$ and $q_0$ are uncertain one can safely say that $-5<q_0<5$
and $0<\Omega_0 < 4 $, hence  $|\Lambda|=21 H_0^2 $.
By setting $H_0=100\rm kms^{-1}Mpc^{-1} $, this leads to $\Lambda\sim 10^{-54} \rm cm^{-2}$.
Now in the Newtonian gravitational theory if $r$ is the distance of a point
mass with respect to the centre of spherically symmetric distribution of matter,
 then the force on this unit mass is given by
 \be
F=-\fr{4\pi}{3}G\rh+\fr{\Lambda}{3}r
\ee
where $\Lambda $ is the Newtonian form of the cosmological constant:\\
(i) $\Lambda > 0$ implies a repulsive force and\\
(ii) $\Lambda < 0$ implies an attractive force.\\
The matter distribution ceases to be a bound system if $F$ is an outward force.
This implies [32] that
\be
\Lambda < 4\pi G\rh
\ee
which gives $\Lambda \sim 10^{-48} \rm cm^{-2}$.
From the above we see that $\Lambda $ acts at dimensions of order of galaxies and has
no effect on the solar system.
\se{De Sitter model}
Consider a flat RW universe with $\rh(t)=\rh_0$=constant.
Equation (4.2) becomes
\be
\fr{\dt R}{R}=H=\rm const.
\ee
Its solution is
\be
R(t)\propto \exp Ht
\ee
This is known as an inflationary solution.
De Sitter obtained this solution for $p=\rh=0$.
However, such a solution is also possible even without $\Lambda$ but with $\rh=\rm const.$
The de Sitter universe is characterized by motion without matter in contrast to the
Einstein's universe which is matter without motion.

One gets  a similar form for $R(t)$ as in the steady-state theory of Bondi and Gold (1948) and Hoyle.
However, unlike in the  de Sitter model, which is empty, in the steady state theory
there is continuous creation of matter due to the so-called c-field [33].
 Although we started with $k=0$ space-time; inflation is not restricted to $k=0$.
For a Universe with $k=+1$ one finds the solution [39]
\be
R\propto H^{-1}\cosh Ht\ ,
\ee
and
for $k=-1$
\be
R(t)\propto H^{-1}\sinh Ht\ ,
\ee
which for large $t$ approaches the inflationary solution.
\section{Importance of the cosmological constant}
There is no convincing evidence available for a nonzero value to the
cosmological constant.
The present interest in the flat cosmological constant models has also appeared
motivated by two reasons:

(i) a $\Lambda$ term helps to reconcile inflation with observations.

(ii) with a $\Lambda$ term it is possible to obtain, for flat universes, a theoretical
age in the observed range, even for a high value of Hubble's constant [34].
\subsection{A varying cosmological term cosmology}
The present estimates of the value of $\Lambda$ are very small. One way to resolve
this dilemma with observations is to assume that $\Lambda$ is not a `pure' constant, but
rather decreases continuously with cosmic expansion. Hence, we can say
$\Lambda$ is extremely small because the Universe is old. Cosmologists postulate
a phenomenological law for the decay of this term [26,12].

Recently (Waga \& Torres), considering the statistics of gravitational lensing,
which is  a powerful tool in constraining models of the Universe, especially
those with a $\Lambda$-term, have shown that cosmologies with a varying
cosmological term give a lower lensing rate.
This is due to the fact that in a varying $\Lambda$ cosmology
the distance to an object with red-shift $z$ is smaller than the distance to the
same object in a constant $\Lambda$ model with the same $\Omega_{0}$. So, the
probability that light coming from the object is affected by a foreground galaxy
is reduced in a decaying $\Lam$ cosmology $[37]$.

Chen and Wu [36] advocated the possibility of particular
$R$-dependence behaviour, i.e. $\Lam\propto R^{-2}$. They argued
in favor of this behaviour of $\Lam$ from some very general
arguments in line with quantum cosmology. From dimensional
consideration and in the sprit of quantum cosmology, one can
always write $\Lam$ as $M_{\rm Pl}^4$ times a dimensionless
product of quantities. For an ansatz for the evolutional behaviour
of $\Lam$, as in the common practice in quantum cosmology, it is
more convenient to use the scale factor $R$ instead of the age of
the Universe. Supposing that no other parameters are relevant
here, the natural ansatz is that $\Lam$ varies according to a
power law in $R$. Theoretically, it can be obtained from some
simple and general assumptions in line with quantum cosmology.
Observationally, it is not in conflict with present data and may
alleviate some problems in reconciling data with the inflationary
scenario.

Therefore, one can write
$$ \Lam\propto M^4_{pl}(\fr{r_{pl}}{R})^n\ ,$$
where $M_{pl}$ and $r_{pl}$ are the Planck mass and length respectively.
With $n=2$ in the above equation, both $\hbar$ and $G$ disappear, since
general relativity is a classical theory, and therefore we
have $ \Lam=\fr{\gam}{R^2}$, where $\gam$ is a dimensionless number of order unity.
The case $n=3$ or $n=1$ would either lead to too big or too small value for
the present cosmological constant $\Lam_0$.

On the same dimensional grounds Carvalho {\it et al.} [42] assumed
$$\Lam\propto\fr{1}{l_p^2}(\fr{t_{pl}}{t_H})^2\ ,$$
where $l_p, t_p $ are respectively the Planck length and time, $n$ is an integer
and $t_H\simeq H^{-1}$ is the Hubble time. Recalling the general relativity is a
classical theory, in order to get rid of $\hbar$ dependence of $n$ one needs
to put $n=2$. Therefore, $\Lam\sim H^2$.\\
On the other hand \"{O}zer and Taha [26] proposed a cosmological model
with $\Lam=\fr{1}{R^2}\ \, $ (i.e. $\gam=1$) based on the critical
energy density assumption.\\
Also a cosmological term varying as $R^{-n},9/5 < n < 2$, was first
introduced by Gasperini [35].


Consider the Friedmann equation
$$ H^2=\fr{8\pi G}{3}\rh-\fr{k}{R^2}+\fr{\Lam}{3}\ .$$
For $k=0, \Lam=3\bt H^2, \bt=\rm const.$, the above equation can be written as
$$ \bt=\fr{\rh_v}{\rh_v+\rh}=\rm const.\ , $$
where $\rh_v=\fr{\Lam}{8\pi G}$ is the vacuum density.
This equation is nothing but the Freese {\it et al.} ansatz. Hence the postulate
$\Lam=3\bt H^2$ is equivalent to the Freese {\it et al.} one.
One can also write the above Friedmann equation  as
$$\Omega+\fr{\Lam}{3H^2}=1\ ,$$
where $\Omega$ is the density parameter.
One can assume that the role of inflation can be described by specifying that
it drives the Universe to a state of geometrical flatness, corresponding to
$\Omega+\fr{\Lam}{3H^2}=1$. It is useful to regard the quantity on the left hand side
of the latter equation as $\Omega_{eff}=1$, with the term $\fr{\Lam}{3H^2}$ regarded as the vacuum
energy contribution to the density parameter.
Using this definition, we see that inflation always drives the Universe
to $\Omega_{eff}=1$.\\
In this work we will investigate some of these laws and discuss their implications on
the evolution of the Universe.
\se{Variable $G$ models}
\subsection{Theories in which $G$ varies with time}
Theories of this type were first proposed by Milne, Dirac, and Jordan.
Later, Brans and Dicke (BD), Hoyle and Narlikar, and Dirac put forward
more elaborate theories of this type.
A variation of $G$ with time has a considerable effect on the evolution of the
Earth and Sun, and on the orbits of the Moon and planets.
If gravity has changed appreciably over the life time of the Earth, the radius of
the Earth might have been affected. It has been suggested that the continents all
fitted together at one time on a much smaller Earth. As the gravitational constant
reduced, the Earth expanded to its present size and the continents were forced
apart. Also a star, like the Sun, in its hydrogen burning phase [3] has
a luminosity $(L)$
\be
L\propto G^{7.7} ,
\ee
as so would have been brighter in the past if $G$ decreased with time $t$.
The effect of this on life on Earth would be enhanced by the fact that the
Earth must be moving away from the Sun if $G$ is declining.
A too fast decline in $G$ would lead to the Sun having already become a red giant.
 A varying $G$ leads to variation in the Moon's distance and period.
 The orbits of the planets are also modified, and this could show up in the radar
 time-delay experiments [24]. It should be emphasized that some theories in which $G$
 varies also predict other changes (to preserve energy conservation, e.g.) which
 can mask the above effects. A  stronger limit on $\dt G$ follows if $^4He$ and $^2H$
 are believed to be synthesized in the big-bang hot phase.
 One then has
\be
|\fr{\dt G}{G}|\le 10^{-12}\rm yr.^{-1}
\ee
The BD cosmology represents the simplest extension of general relativity (GR).
In addition to the tensor gravitational fields represented by the metric tensor,
there is a scalar field (related to the gravitational `constant') which is a function of time.
BD take $\Lambda=0$ and seek to satisfy Mach's principle, that local inertia
properties should be determined by the gravitational field of  the rest matter
in the Universe, by taking
\be
G^{-1}\sim \sum_{Universe}\fr{m}{r}
\ee
The models for $k=0$ are particularly simple, since, there
$$ R\propto t^q, \ \  \ G\propto t^r$$
where $$ q=\fr{2(1+w)}{4+3w}, \ \  r=\fr{2}{4+3w}$$
 $w$ being  the `coupling constant' between the scalar field and geometry:
$w\rightarrow \infty $ gives the Einstein-de Sitter model.
Note that for general $w,\ G\rh t^2=\rm const.$.
Dirac's 1937 model is obtained by setting $w=-2/3$. However, BD theory makes
slightly different predictions from GR for the deflection of light by
the Sun and for the advance of planets [25].
And if helium is synthesized in the big bang, there is an even stronger
limit, \ $w > 100$ [24].
Also if $G$ has had greater values in the past than it has now, one would
expect a small mass density to have the same effects as a larger mass density
has later. An increasing $G$ would lead to a contraction of the earth.
\se{A new formalism of variable $G$ models }
Ever since Dirac first considered the possibility of variable gravitational
``constant", $G$ [4], there have been numerous modification of general relativity
to allow for variable $G$ [5]. These theories have not gained wide acceptance.
However, recently [6-10] a modification has been proposed treating $G$ and the
cosmological `constant' $\Lambda$ as non constant coupling scalars.
Einstein's equations
\be
{\cal{R}}_{\mu\nu} - \fr{1}{2}{\cal{R}} g_{\mu\nu} = 8\pi G T_{\mu\nu} + \Lambda g_{\mu\nu},
\ee
are considered, where $G$ and $\Lambda$ are coupling scalars and the other symbols
have their usual meanings .
The principle of equivalence then demands that only $g_{\mu\nu}$ and not $G$ and
$\Lambda$, must enter the equation of motions of particles and photons.
This approach is appealing because it leaves Einstein's equation formally un changed
since variation in $\Lambda $ is  accompanied by a variation in $G$ [6-10],
in such a way the usual energy conservation law
$ T^{\mu\nu}_{;\nu} = 0,$ holds.
If we take the divergence of (5.4) and use the Bianchi identities, we get
\be
G_{;\nu} T^{\nu\mu} + \Lambda_{;\nu} g^{\nu\mu} = 0
\ee
This approach, however, is non covariant and the field equations can not be
derived from a Hamiltonian. We note that the propagation equations for the scalars
are not contained in eqs.(4.24).
Despite this drawback, there are several advantages of the present approach.
This approach could be a limit of a more viable fully covariant theory such as five dimensional
Kaluza-Klein theory [11].
Various possibilities for variable $G$ can be investigated. The problems of the
standard model could be solved as in the inflationary scenario [12].
The variation of $G$ with time is reasonable, since $G$ couples geometry to matter ,
and in an expanding Universe we expect $G = G(t)$.
It is reasonable to assume that the Universe had always the Einstein-de Sitter
critical density.
For unless the Universe had this, it would have diverged, very rapidly from it.
With the present Hubble constant $H_{p}= 5\times 10^{-11}\rm yr^{-1}$, the present
critical energy density $\rh=\fr{3H^2_p}{8 \pi G}\simeq 2\times 10^{-47} \rm GeV^4$.
The current energy density of the Universe, on the other hand, is between $10^{-47}\rm GeV^4$
and $10^{-48}\rm GeV^4 [26]$.

With these assumptions we will show that there are models in which $G$ increases
with time and other models in which it decreases.\\
Recently, the flat Friedmann Robertson  Walker (FRW) models have been studied
with the present formalism. A number of solutions were presented including de
Sitter-type ones relevant to inflation [30,31].
\subsection{Bulk viscous solutions}
The role of viscosity in cosmology has been studied by various authors [18,19,29].
It was initially hoped that neutrino viscosity could smooth all initial anisotropies
and leads to the isotropic Universe that we observe today [13-15,22].
The bulk viscosity associated with the Grand Unified Theory (GUT) [6,16,17] phase
transition can lead to an inflationary universe.
It is also known that the introduction of bulk viscosity  can avoid the big bang
singularities [18,19,27]. When viscosity is introduced in a fluid the fluid
becomes imperfect, i.e. the one in which the pressure, density and velocity
vary appreciably over distances of the order of a mean free path, or over
times of the order of mean free time, or both. For such fluids, the kinetic
energy is dissipated as heat. For relativistic fluids, the dissipative effects
play an important role in the history of the early Universe. The energy -
momentum tensor of the imperfect fluid $(T'_{\alpha\beta})$ takes the form [23]
\be
T'_{\alpha\bt}=T_{\alpha\bt}+\Delta T_{\alpha\bt}
\ee
where ($\Delta T_{\alpha\bt}$) is regarded as a correction term to the energy
momentum-tensor of the perfect fluid $T_{\alpha\bt}$.
In a comoving frame $\Delta T_{00}=0$, and in a general frame it satisfies
\be
u^\alpha u^\beta\Delta T_{\alpha\bt}=0
\ee
The most general form containing viscosity allowed by this condition and the
second law of thermodynamics is (see Appendix A)
\be
\Delta T_{\alpha\bt}=\zeta
H^{\alpha\gamma}H^{\bt\delta}W_{\gamma\delta}+\eta
H^{\alpha\bt}u^\gam_{;\gam} \ee where
$W_{\alpha\bt}=u_{\alpha;\bt}+u_{\bt;\alpha}-\fr{2}{3}g_{\alpha\bt}u^\gamma_{;\gamma}$
is the shear tensor and $H_{\alpha\bt}=g_{\alpha\bt}-u_\alpha
u_\bt$ is the projection tensor on a hyperplane normal to
$u^\alpha$. $\eta$ and $\zeta$ are the coefficient of bulk and
shear viscosity, respectively. We will here consider the bulk
viscosity only since the shear viscosity plays no role in a RW
model. Therefore $$ \Delta
T_{\alpha\bt}=\eta(g_{\alpha\bt}-u_\alpha
u_\bt)u^\gamma_{;\gamma}$$ Hence $$ T'_{\alpha\bt}=(\rh+p-\eta
u^\gamma_{;\gamma})u_\alpha u_\bt-(p-\eta
u^\gamma_{;\gamma})g_{\alpha\bt}$$ If we now replace $p-\eta
u^\gamma_{;\gamma}$ by $p^*$\footnote{The total pressure $p^*$
accounts for the isotropic pressure $p$ plus viscous terms can be
represented as a polynomial in $\theta\  \
(\theta=u^\gam_{;\gam}):\ \ \ p^*=p-\sum_{k=1}^N\alf_k\theta^k$
where $\alf_k$ are in general functions of \ $\rh$ [21,28,41].} ,
the above equation becomes $$ T'_{\alpha\bt}=(\rh+p^*)u_\alpha
u_\bt-p^*g_{\alpha\bt}$$ Apart from the above replacement, this
equation looks the same as that of a perfect fluid. Therefore, the
introduction of the bulk viscosity does not alter the isotropy and
homogeneity of the Universe.

The field equations with bulk viscosity can be obtained from the general relativistic field
equations by replacing the pressure term, $p$, by, $p^*$, where
\be
p^* = p -\eta u^\gam_{;\gam}
\ee
In a Robertson-Walker model, we have $u^\gam_{;\gam}=3H$, so that $p^*=p-3\eta H$.

In eq.(4.27) $\eta$ is usually taken to have a power law form [19]
\footnote{In isotropic and homogeneous models all parameters
depend only on the time $t$, and therefore we may consider them as
functions only of the energy density \ \ $\rh$.}
\be
\eta =\eta_0 \rh^n
\ee
Where $\eta_0 \ge 0 $, $ \rh$ is the energy density, $n$ is a constant and  $H$ is
the Hubble constant.
In addition to the linear dependence of $\eta$ upon $H$, as in the above equation,
some workers [20,21,28] have considered a quadratic dependence upon $H$, i.e.
\be
p^* = p - 9 \zeta H^2
\ee
The $\zeta=\rm const $ models were analyzed by Romero [20] .
The bulk viscous models considered so far are endowed with particle creation .
In the Chapter 7 and 8 we will consider a model with variable $G$ and $\Lambda$
and bulk viscosity .
This combination of $G$, $\Lambda$ and $\eta$ has not been considered before.
The  model turns out to have many interesting features. Various models could be
reproduced from this model by taking  particular values of $n$, where $\eta =\eta_0\rh^n$.
\se{References}
$[1]$ R.W.Hellings, P.J.Adams, J.D.Anderson, M.S.Keesey, E.L.Lau and E.M.Dtandish
{\it Phys.Rev.Lett.51,18,1609}(1983)\\
$[2]$ T.L.Chow, {\it Nouvo Cimento Lett.31,4,119}(1981)\\
$[3]$ L.S.Levitt, {\it Nouvo Cimento Lett.29,23}(1980)\\
$[4]$ P.A.M.Dirac, {\it Nature 139, 323}(1937)\\
$[5]$ P.S.Wesson, Gravity particles and Astrophysics (1980, Reidel, Dordrecht)\\
$[6]$ Y.K.Lau,{\it Austr.J.Phys.39, 339}(1985)\\
$[7]$ A.Beesham, {\it Nouvo Cimento 96B, 179}(1986)\\
$[8]$ A.Beesham, {\it Inter.J.Theor.Phys.25,1295}(1986)\\
$[9]$ A.-M.M.Abdel Rahman, {\it Gen.Rel.Gravit. 22, 655}(1990)\\
$[10]$ M.S.Berman, {\it Phy.Rev.D43, 1075}(1991)\\
$[11]$ P.S.Wesson, {\it Gen.Rel.Gravit.16, 193}(1984)\\
$[12]$ M.S.Berman, {\it Gen.Rel.Gravit. 23,465}(1991)\\
$[13]$ J.M.Stewart, {\it Astrophys. Lett.2,133}(1968)\\
$[14]$ A.G.Doroshkevich, Ya.B.Zeldovich and I.D.Novikov, {\it Zh.Eksp.Teor.Fiz.53,644}(1967)\\
$[15]$ C.B.Collins and J.M.Stewart, {\it Mon.Not.R.Astron.Soc.153,419}(1917),\\
$[16]$ I.Waga, R.C.Falcao, and R.Chandra, {\it Phy.Rev.D33, 1839}(1986)\\
$[17]$ T.Pacher, J.A.Stein-Schabes, and M.S.Turner, {\it Phy.Rev.D36, 1603}(1987)\\
$[18]$ G.L.Murphy, {\it Phy.Rev.D8, 4231}(1973)\\
$[19]$ V.A.Belinkskii and I.M.Khalatinkov, {\it JETP Lett.99}(1975)\\
$[20]$ C.Romero, {\it Rev.Bras.Fis.18,75}(1988)\\
$[21]$ M.Novello and J.B.S.d'Olival, {\it Acta Phy.Pol.B11,3}(1980)\\
$[22]$ C.W.Misner, {\it Nature(London)214,40}(1967)\\
$[23]$ S.Weinberg, Gravitation and Cosmology (Wiley, New York, p.593 1972)\\
$[24]$ Row-Robertson, Michael, Cosmology (Oxford, Clarendon Press, 1981).\\
$[25]$ C. Brans and R.H. Dicke, {\it Phy. Rev. D124, 203}(1961)  \\
$[26]$ M. \"{O}zer and M.O.Taha, {\it Nuc. Phy. B287, 776}(1987)    \\
$[27]$ Z.Golda, M.Heller, and M.Szydlowski, {\it Astrophys. Space.Sci. 90, 313}(1983) \\
$[28]$ M.Novello and R. A. Araujo,  {\it Phys. Rev.D22, 260}(1980)\\
$[29]$ O.Gron, {\it Astrophys. Space Sci.173,191}(1990)  \\
$[30]$ D. Kalligas, P. Wesson, and C.W.Everitt, {\it Gen. Rel. Gravit. 24, 351}(1992) \\
$[31]$ C.Wolf, {\it S.-Afr.Tydskr Fis.14, 68}(1991)   \\
$[32]$ J.A.Islam, Introduction to Mathematical Cosmology (Cambridge University Press, 191)\\
$[33]$ F.Hoyle and J.V. Narlikar, {\it Proc. Roy. Soc. A270, 334 }(1962)\\
$[34]$ S. M. Caroll, W.H. Press and E.L. Turner, {\it Ann. Rev. Astron. Astrophys. 30, 499}(1992)\\
$[35]$ M. Gasperini, {\it Phy. Lett.B194, 347}(1987)\\
$[36]$ W. Chen and Y.Wu, {i\ Phy. Rev.41, 695}(1990)\\
$[37]$ Creation of the Universe, Li. F. Zhi, 1989 (World Scientific)\\
$[38]$ Introduction to Cosmology, J. V. Narlikar, 1983 (Jones \& Bartlett)\\
$[39]$ Introduction to Cosmology, M. Roos, 1994 (John Wiley)\\
$[40]$ The Isotropic Universe, D. J. Raine, 1981 (Adam Hilger)\\
$[41]$ J.A.S. Lima, R.Portugal and  I. Waga {\it Phys. Rev.D37}2755(1988)\\
$[42]$ J. C. Carvalho, J. A. S. Lima and I. Waga, Phys. Rev. D. 46 (1992) 2404
\chapter{A Vacuum Decaying Cosmological Model}
\se{Introduction}
The cosmological constant problem $[1]$, that phase transitions in the
 early Universe would have left the cosmological constant larger than
 the observed upper bound, $\Lambda\ge 10^{-84} (\rm GeV)^2 [1,2]$, by about 120
 order of magnitude, is a major puzzle of cosmology and particle physics.
 Considerable efforts were made in seeking its solution $[1]$.
 One approach attempts to avoid this impass$\acute{e}$ by allowing $\Lambda$ to vary smoothly
 with time so that models in which it was appreciable in the past could be
 constructed $[1]$. Examples are the models of \"{O}zer and Taha $[1]$, Chen and Wu
 $[4]$, as well as some of their later generalizations $[5,6]$, all of which
 require $\Lambda\propto R^{-2}$, where $R$ is the Robertson-Walker (RW) scale factor.
 Recently Carvalho, Lima and Waga $[7]$ proposed the ansatz
 \be
 \Lambda= 3\bt H^2+\fr{3\gamma}{R^2}
 \ee
 where $\beta$ and $\gamma$  are dimensionless numbers( natural units being used)
 and $H=\fr{\dot R}{R} $ is Hubble's constant ( an overdot denotes time
  differentiation) . They suggested the $\beta$ -term in Equation (5.1) on the basis
  of simple dimensional arguments consistent with quantum  gravity. Special cases
  of eq.(5.1) when $\bt $=0 are :
Chen and Wu $[4]$ singular model ($\gamma$ arbitrary ), \"{O}zer - Taha (OT) $[3]$
nonsingular cosmology ($\gamma =1$), and the singularity - free models of Ref.  $[6]$
($1/2 <\gamma <1$).
Carvalho {\it et al} $[7]$ have investigated the effect of $\bt$-term in eq.(5.1)
 on the singular model of Chen and Wu $[4]$. In the present work we study the
 implications of this term for nonsingular cosmologies of \"{O}zer - Taha
 $[3]$ type.\\
\se{Field equations}
 In a RW universe with a perfect fluid energy - momentum tensor, Einstein's equations
 with a variable $\Lambda$ give ($\alf\equiv 3/8\pi G):$
 \be
 \alf^{-1}\rh= [ \fr{\dt R}{R}]^2+\fr{k}{R^2}-\fr{\Lambda(t)}{3}\ \ ,
 \ee
 \be
 \fr{3}{2}\alf^{-1}(\rh+p)=[ \fr{\dt R}{R}]^2 +\fr{k}{R^2}-\fr{\dd R}{R}\ .
 \ee
 In these equations $\rh $ and $p$ are the cosmic energy density and pressure and
$k$ the curvature index.
\se{ Nonsingular Model  }
The introduction of eq.(5.1) and the radiation equation of state p=$\fr{1}{3}\rh$
in eqs.(5.2) and (5.3) lead to ($\bt \neq \fr{1}{2} $):
\be
\dt R^2 = \fr{(2\gamma -k)}{(1 - 2\bt)}+A_0R^{-2+4\bt}\ ,
\ee
\be
\rh=\fr{\alf(\gamma-\bt k)}{(1-2\bt)}R^{-2}+\alf(1-\bt)A_0 R^{-4+4\bt},
\ee
\be
\rh_v=\fr{\alf(\gamma-\bt k)}{(1-2\bt)} R^{-2}+\alf\bt A_0 R^{-4+4\bt},
\ee
Where $A_0$ is a constant and $\rh_v =\fr{\alf}{3}\Lambda$ is the vacuum energy
density.
Equations (5.4)-(5.6) were analyzed by Carvalho {\it et al} $[7]$ for the case $A_0>0$
and $ \bt < 1$ corresponding to a singular universe. As pointed by them $[7]$ a
"natural extension (of their work ) would be to explore different scenarios
obtained by making $A_0 < 0  $," for which the cosmology is nonsingular.
One possible line for such an investigation would be to consider a nonsingular
cosmology of the OT type based on eq.(5.1). This is the theme of what follows.
\se{Radiation Universe}
Requiring $A_0 < 0$ in eq.(5.4) implies
\be
\fr{2\gamma-k}{1-2\bt} > 0.
\ee
It is then possible for R to have had an initial minimum non vanishing value at
t=0, say. The necessary condition for the existence of this minimum in an expanding
Universe is $\dt R=0$ at t=0. We explore this possibility.
In order to reach $\dt R=0$ at t=0 the exponent of $R$ in eq.(5.4) must be negative,
implying that $\bt < 1/2 $. Hence from eq.(5.7),
\be
k < 2\gamma.
\ee
Let R=$R_0$ at t=0, when $\dt R=0$. Clearly $ R_0\neq 0$ or else $\dt R^2(0)=-\infty$.
Thus
\be
\dt R^2=\fr{(2\gamma-k)}{(1-2\bt)}[1-\fr{R_0^{2-4\bt}}{R^{2-4\bt}}] < \fr{2\gamma-k}{1-2\bt},
\ee
\be
\rh=\fr{\alf(\gamma-\bt k)}{(1-2\bt)R^2}[1-\fr{(2\gamma-k)(1-\bt) R_0^{2-4\bt}}{(\gamma-\bt k)R^{2-4\bt}}],
\ee
\be
\rh_v=\fr{\alf(\gamma-\bt k)}{(1-2\bt)R^2}[1-\fr{\bt(2\gamma-k)R_0^{2-4\bt}}{(\gamma -\bt k)R^{2-4\bt}}].
\ee
From eq.(5.10),
\be
\rh_0=\fr{\alf(k-\gamma)}{R_0^2}
\ee
which is the same as eq.(5.9) of Ref.$[6]$. Thus the physical condition $\rh_0\geq 0$
requires.
\be
k\geq\gamma.
\ee
Equations (5.8) and (5.12) lead to
\be
1/2 < \gamma \le k = 1,
\ee
implying a closed universe. The result (5.14) coincides with eq.(5.13) of  Ref. [6].
Classical general relativity without or with constant $\Lambda$ does not explain
the origin of cosmic entropy because classical Einstein's equations are purely
adiabatic and reversible. In the present cosmology, however, the change in the
entropy $S$ is related to the temperature $T$ and the change in $\rh_v$, by [6]
\be
TdS = - R^3d\rh_v ,
\ee
so that from eq.(5.11),
\be
T\fr{dS}{dR}=\fr{2\alf(\gamma-\bt)}{1-2\bt}[1-\fr{2\bt(2\gamma-1)(1-\bt)R_0^{2-4\bt}}{(\gamma-\bt)R^{2-4\bt}}],
\ee
in the radiation Universe. Therefore $\fr{dS}{dR}> 0 $ provided the second term
inside the bracket of eq.(5.16) is less than unity. Since $2\bt < 1, 0 < (2\gamma - 1 \le 1)$
but $\fr{1-\bt}{\gamma-\bt}\ge 1$, this would be guaranteed, for $R\ge R_0$, if we
simply choose $\gamma = 1$. Henceforth we restrict ourselves to the model with $\gamma = 1, \bt\neq 0$.
We will consider two cases separately
\se{Nonsingular model: case 1.}

Putting $\gamma = k =1$ in eqs.(5.9)-(5.12) we obtain
\be
\dt R^2 = \fr{1}{1-2\bt}[1-\fr{R_0^{2-4\bt}}{R^{2-4\bt}}] < (1-2\bt)^{-1},
\ee
\be
\rh = \alf(1-\bt)H^2,
\ee
\be
\rh_v = \fr{\alf(1-\bt)}{(1-2\bt)R^2}[1-\fr{\bt R_0^{2-4\bt}}{(1-\bt)R^{2-4\bt}}],
\ee
\be
\rh_0 = 0.
\ee
these equations extend the $\gamma = 1, \bt = 0$ OT model, albeit in a different direction
than that of Ref.[6]. As in the OT model the initial Universe  here also is empty
(and cold) but the density parameter
\be
\Omega\equiv \fr{\rh}{\alf H^2} = 1-\bt < 1 ,
\ee
compared to $\Omega \equiv 1$ in the OT model. The density $\rh$ attains a maximum
\be
\rh_{max} = \fr{\alf}{2R_0^2[2(1-\bt)]^{\fr{1}{1-2\bt}}} < \fr{\alf}{2R_0^2}\ ,
\ee
at
\be
R = R_{max} = [2(1-\bt)]^\fr{1}{2-4\bt}R_0 > R_0 .
\ee
Note that in Ref.[6]  $\rh_0 < \alf/2R_0^2$ and $\rh_{max}\ge\alf/4R_0^2.$
An estimate of $R_0$ and hence an upper bound on $\rh_{max}$ can be deduced
as follows. From eq.(5.17),
\be
0< \dd R \le R_0^{-1}\ ,
\ee
implying the existence of a natural cosmic acceleration limit in the radiation
Universe. Such a maximal acceleration, of the order of Planck mass $M_{pl}=G^{-1/2}$,
($h/2\pi=c=1)$, has been discussed before \ [8]. Thus taking $R^{-1}_0\sim M_{pl} $ yield
$ R_0\sim G^{1/2}\approx 8\times 10^{-20} (\rm GeV)^{-1} = 1.6\times 10^{-33} \rm cm$,
so that $\rh_{max}  < 3\times 10^{95}\rm kgm^{-3}$.
Finally, for $R \gg R_0, R\sim t $(linear expansion) and $ \rh=\rh_v\sim t^{-2}$.
The variation $\rh_v\propto t^{-2} $ is regarded by Berman\ [9] as more
fundamental that expression of $\rh_v$ in terms of $R$, e.g.
$\rh_v\propto R^{-2}$ \ [3-6] .
\se{Radiation and matter}
Equation (5.2) and (5.3) may be combined to give
\be
\fr{dE}{dR}+3pR^2=-\fr{\alf}{3} R^3\fr{d\Lambda}{dR}\ \ ,
\ee
where $E=\rh R^3$ and $\Lambda $ may be written as ($\gamma=k=1)$:
\be
\Lambda=\fr{3\alf^{-1}\bt\rh}{(1-\bt)}+\fr{3}{R^2}. \ee These
equations are assumed to hold throughout cosmic evolution.
Following \"{O}zer and Taha \ [3] and Ref. [6] we assume the
Universe to have evolved through: a very early era, $R_0\le R\le
R_1$, say, of pure radiation (discussed in subsection ); a
subsequent period of $R_1\le R\le R_2$ of  matter generation; and,
lastly, for $R\ge R_2$, an era of radiation and conserved non
relativistic matter reaching to the present. Except, perhaps, for
the matter generation epoch the vacuum is assumed to decay into
radiation only. Hence with the radiation and matter densities
denoted by $\rh_r$ and $\rh_m$ respectively ($\rh_r+\rh_m=\rh$),
the energy, in volume $R^3$, of non relativistic matter is $
E=\rh_m R^3=\rh_{mp}R_p^3=E_{mp}$, where the subscript ``p"
designates present-day quantities. We also assume that non
relativistic matter has zero pressure so that $ p =
p_r=\fr{1}{3}\rh_r$. Chen and Wu  \ [4] and also Carvalho {\it et
al} \ [7] assume that the vacuum decays into non relativistic
matter in the matter-dominated (MD) universe. But Freese {\it et
al} \ [10] have demonstrated that such a scenario is not,
apparently, favored by observations. Under our assumptions
eqs.(5.25) and (5.26)  yield, when $R\ge R_2$ ,
\be
\fr{d\rh_r}{dR}+\fr{4(1-\bt)}{R}\rh_r=\fr{3\bt\rh_{mp} R_p^3}{R^4}+\fr{2\alf(1-\bt)}{R^3}\ .
\ee
This has the solution
\be
\rh_r=\fr{3\bt\rh_{mp} R_p^3}{(1-4\bt)R^3}+\fr{\alf(1-\bt)}{(1-2\bt)R^2}[1+\fr{\om R_p^{2-4\bt}}{R^{2-4\bt}}]\ \ ,
\ee
where

$$(1+\om)=\fr{(1-2\bt) R_p^2\rh_{mp}}{\alf(1-\bt)}[\fr{\rh_{rp}}{\rh_{mp}}-\fr{3\bt}{1-4\bt}]$$
\be
=(1-2\bt)[\rh_{rp}-\fr{3\bt}{1-4\bt}\rh_{mp}][(1-\bt)\rh_{vp}-\bt(\rh_{rp}+\rh_{mp})]^{-1}.
\ee
Hence from eq.(5.26),
\be
\rh_v=\fr{\bt\rh_{mp}R_p^3}{(1-4\bt)R^3}+\fr{\alf(1-\bt)}{(1-2\bt)R^2}[1+\fr{\bt\om R^{2-4\bt}_p}{(1-\bt)R^{2-4\bt}}],
\ee
and by eq.(5.2),
\be
\dt R^2=\fr{\alpha^{-1}\rh}{(1-\bt)}R^2\\
=\fr{\alpha^{-1}\rh_{mp}R^3_p}{(1-4\bt)R}+\fr{\om R_p^{2-4\bt}}{(1-2\bt)R^{2-4\bt}}+(1-2\bt)^{-1}.
\ee
From eq.(5.31),
\be
\Omega=1-\bt < 1 ,
\ee
as before (see eq.(5.21)). Lastly, the present value of the deceleration parameter
$q= -R\dd R/\dt R^2 $ is
\be
q_p=[1/2+\fr{\om(1-4\bt)(1-2\bt)}{(1+\om)(1-\bt)}[\fr{\rh_{rp}}{\rh_{mp}}-\fr{3\bt}{1-4\bt}]]\times [1+\fr{1-4\bt}{1-\bt}[\fr{\rh_{rp}}{\rh_{mp}}-\fr{3\bt}{1-4\bt}]]^{-1}.
\ee
For $\bt=0 $ eqs.(5.28)-(5.33) reduce to the corresponding OT results.
Denote by $R_{eq}$ the  value of $R$ at $t=t_{eq}$, the time when radiation
and matter were balanced in equilibrium. Let $E_r=\rh_r R^3$, where $\rh_r$ is
given by eq.(5.28), be the radiation energy. Then
\be
E_r(R_{eq})=E_{mp}
\ee
\be
\begin{array}{c}
E_r(R) > E_{mp} \ \ \  \ \rm when\ \  R< R_{eq}\\
E_r(R) < E_{mp}\ \ \ \ \ \rm  when\ \  R>R_{eq}
\end{array}
\ee
The condition that $E_r$ was decreasing as R approached $R_{eq}$ would then
imply that
\be
(1-4\bt)\om > \fr{R_{eq}^{2-4\bt}}{R_p^{2-4\bt}}.
\ee
Hence either
\be
\om > 0\ \ \ and \ \bt < 1/4 ,\\
\rm or,\\
\om < 0 \ \ \ and \  \ 1/4 <\bt < 1/2.
\ee
We discuss each case separately .
\bc
$1. \ \ \om > 0, \ \ \ \bt < 1/4$
\ec
From eq.(5.29) we have
\be
\fr{3\bt}{1-4\bt} < \fr{\rh_{rp}}{\rh_{mp}}\equiv\delta_{rp}\ .
\ee
Hence
\be
\bt < \delta_{rp}.
\ee
There are three subcases:\\
(a) $\bt = 0 $: This corresponds to the OT \ [3] model.\\
(b) $\bt > 0$ :  Assuming $\delta_{rp} \ll 1 $, i.e. the present Universe
is MD, we have $0 < \bt \ll 1 $. Also from eq.(5.30) $d\rh_v/dR < 0$, so that
entropy is continuously generated. From  eq.(5.28),
\be
\delta_r\equiv\fr{\rh_r}{\rh_m}=\fr{3\bt}{1-4\bt}+\fr{\alf(1-\bt)R}{(1-2\bt)\rh_{mp}R^3_p}\times[1+\fr{\om R_p^{2-4\bt}}{R^{2-4\bt}}].
\ee
In the early radiation and matter universe $\delta_r\gg 1$. Using this condition
in eq.(5.40) and noting from (5.38) that\\
$3\bt(1-4\bt)^{-1} < \delta_{rp}\ll 1$ one has
\be
1+\fr{\om R_p^{2-4\bt}}{R^{2-4\bt}} \gg \fr{(1-2\bt)\rh_{mp}R^3_p}{\alf(1-\bt) R}.
\ee
But by condition (5.36)
\be
\fr{\om R_p^{2-4\bt}}{R^{2-4\bt}} > \fr{R_{eq}^{2-4\bt}}{(1-4\bt)R^{2-4\bt}}\gg 1,
\ee
for $R_{eq}/R \gg 1 $ and $0 < \bt < \delta_{rp}\ll 1. $ Hence from
eqs.(5.31), (5.41) and (5.42) ,
\be
\dt R^2\approx\fr{\om R_p^{2-4\bt}}{(1-2\bt)R^{2-4\bt}}\approx\fr{\alpha^{-1}\rh_r R^2}{(1-\bt)},
\ee
when $R_{eq}/R \gg 1$.
Consider now $\rh_v$ of eq.(5.30). Assuming, plausibly, that for $ R\ll R_{eq}$
the vacuum energy was much more important than the energy of non relativistic
matter, i.e. $\rh_v/\rh_m\equiv\delta_v\gg 1$, we have
\be
1+\fr{\bt\om R_p^{2-4\bt}}{(1-\bt)R^{2-4\bt}}\gg \fr{(1-2\bt)\rh_{mp}R_p^3}{\alf(1-\bt)R}\ .
\ee
But by (5.36),
\be
\fr{\om\bt R_p^{2-4\bt}}{(1-\bt)R^{2-4\bt}} > \fr{\bt R_{eq}^{2-4\bt}}{(1-\bt)(1-4\bt)R^{2-4\bt}}.
\ee
 For $ R\ll R_{eq} $, and provided $\bt$ is not negligibly small ( much smaller than $\delta_{rp}$)
, the RHS of (45) will be considerably larger that unity. ( If $\bt\ll\delta_{rp}\ll 1$
, the $\bt$-term in $\Lambda $ may be dropped, which is subcase (a)).
 Then eqs.(5.30), (5.45), (5.44) and (5.43) imply
 \be
 \rh_v\approx\fr{\alf\bt\om R_p^{2-4\bt}}{(1-2\bt) R^{4-4\bt}}\approx\fr{\bt\rh_r}{(1-\bt)}.
 \ee
 The relation between $\rh_v$ and $\rh_r$  in this result is identical to eq.(5.7)
 in the paper of Freese {\it et al}. \ [10] where a parameter $x=\rh_v/(\rh_r+\rh_v) $
 replaces $\bt$. But whereas letting $x\rightarrow 0 $ in Ref. \ [10] produces standard
 cosmology, one is not entitled to the limit $\bt \rightarrow 0 $ in  eq.(5.46)
 here because the derivation of this equation assumed that $\bt $ is not
 vanishingly small.\\
 (Taking naively  $\bt\rightarrow 0 $ in eq.(5.46) one concludes erroneously that $\rh_v= 0 $
in the OT model). Integration of eq.(5.43), taking approximately $t=t_2\approx 0$
when $R=R_2\approx 0$, gives
\be
R=[\fr{2(1-\bt)\om^{1/2}R_p^{1-2\bt}}{(1-2\bt)}]^{(1/2(1-\bt)} t^{1/(2(1-\bt)}\ .
\ee
Hence from eq.(5.46),
\be
\rh_v\approx \fr{\alf\bt}{4(1-\bt)^2 t^2}\approx\fr{\bt}{(1-\bt)}\rh_r \ ,
\ee
which is the same as eq.(5.8) of Freese {\it et al}. $ [10]$. As noted by them in
eq.(5.48) has been suggested by various authors $ [11]$, with $\bt$ presumably
dependent on the particular model of vacuum decay. In the present model
$0 < \bt < \delta_p $.\\
(c) $\bt=-\mid\bt\mid < 0$ : Then one has from  condition (5.36),
\be
1-\fr{\mid\bt\mid\om R^{2+4\mid\bt\mid}_p}{(1+\mid\bt\mid)R^{2+|\bt|}}
\ee
$$< 1-\fr{\mid\bt\mid R_{eq}^{2+4\mid\bt\mid}}{(1+\mid\bt\mid)(1+4\mid\bt\mid)R^{2+4\mid\bt\mid}}\ .$$
Thus provided $\mid\bt\mid$ is not vanishingly small both sides of
this inequality will be negative when $R\sim R_2, R_{eq}/R \gg 1$.
(If $\mid\bt\mid$ is vanishingly small subcase (a) is retrieved). It then follows from eq.(5.30) that
$\rh_v < 0$ and $\fr{d\rh_v}{dR} > 0 $ for $ R\sim R_2 $, implying a decreasing
 entropy in the early radiation and matter universe. Hence we exclude $\bt < 0$\\
 Observe that eq.(5.31) implies $\dd R< 0$ if $\om >0 $ and $\bt <1/4 $.
 On the other hand from eq.(5.24) $\dd R> 0$ in the radiation universe.
 This reversal of the sign of $\dd R $ is suggestive  of an intermediate
 phase  transition period separating the pure radiation and the radiation
 and matter eras.
 \bc
 2. $\om = -\mid \om \mid < 0 $ , $ 1/4 < \bt < 1/2  .$
 \ec
 Consider eq.(5.31). Combining the second and third terms on its RHS and using
 condition (5.36) give
 \be
 \fr{1}{(1-2\bt)}[1-\fr{\mid\om\mid R_p^{2-4\bt}}{R^{2-4\bt}}]
 \ee
 $$ < \fr{1}{(1-2\bt}[1-\fr{R_{eq}^{2-4\bt}}{(4\bt-1)R^{2-4\bt}}].$$
 For $ R_2 \le R \le R_{eq} $ the expression to the right of the inequality, and
 so also that to its left, is negative. This leads to $\dt R^2 < 0 $ and $\rh < 0 $
 in eqs.(5.31) and (5.31), which is physically inadmissible. Thus we conclude
 from this subsection that $\om > 0 $ and $ 0 < \bt < \delta_{rp} $, where
 $\delta_{rp} $ is present ratio of radiation-to-matter energy density.
\se{Phase transition}
 From eqs.(5.25) and (5.26) one has
 \be
\fr{d E}{d R}=2\alf(1-\bt)+3\bt R^2\rh +-3(1-\bt)R^2 p  \ ,
\ee
leading, on using $E_r=\rh_r R^3$ and eq.(28), to
$$ 3\int_{R_0}^{R_2}R^2[(1-\bt)p-\bt\rh]dR = 2\alf(1-\bt)(R_2-R_0)-E_2 $$
\be=-2\alf(1-\bt)R_0-\fr{(1-\bt)E_{mp}}{(1-4\bt)} + \fr{\alf(1-\bt)(1-4\bt)R_2F(\om)}{(1-2\bt)},
\ee
where
\be
F(\om)\equiv 1-\fr{\om R^{2-4\bt}_p}{(1-4\bt)R_2^{2-4\bt}}\ \ .
\ee
But by condition (5.36),
\be
F(\om) < 1-\fr{ R^{2-4\bt}_{eq}}{(1-4\bt)^2 R_2^{2-4\bt}} < 0 .
\ee
Hence the integral in eq.(5.52) in negative implying that
$$\fr{3p}{\rh} < 3\bt(1-\bt)^{-1} < 1, \ \ \ (\bt< 1/4)\ ,$$
or by condition (5.38),
\be
\fr{3p}{\rh} < 3\bt(1-\bt)^{-1} <  3\bt(1-4 \bt)^{-1} < \delta_{rp},
\ee
during at least part of the period $(R_1, R_2) $.
The condition (5.55) is clearly incompatible with the pure radiation
 equation of state. The condition (5.55) is incompatible with the
 equation  $\fr{3p}{\rh} =1-\fr{E_m}{E} $ of radiation and matter. For if one were to suppose
 that it is consistent with this equation one would have  $ 1-\fr{E_m}{E}< \delta_{rp}$
, which does not admit simultaneously the assumptions of a present MD universe
 ($\delta_{rp} \ll 1$) and an early radiation - dominated universe ( $E_m/E\ll 1$).

Thus we interpret the era $(R_1, R_2)$ as a phase transition period associated
with the creation  of rest - mass and the appearance of decelerated cosmic
expansion. During this period the pressure may become negative. (If $\bt =0$
negative pressure must occur.)
\se{Baryon-to-photon ratio}
If the radiation produced by vacuum has a Planckian thermal distribution
then its temperature  $T$ will be related to its density $\rh_r$ by
\be
\rh_r=\fr{\pi^2}{30}g_{eff}T^4\ .
\ee
where $g_{eff}$ is the effective number of spin degrees of freedom.
Then in the early ($R\ll R_{eq}$) radiation and matter universe we have
from eq.(5.56) and (5.48),
\be
T(t)\approx [\fr{15\alf}{2\pi^2(1-\bt)}g_{eff}]^{1/4}t^ {-1/2}\ ,
\ee
which together with eq.(5.47) imply
\be
T\sim R^{\bt-1}\ .
\ee
Equations (5.57) and (5.58) are identical to the corresponding results
of Freese {\it et al}. $ [10]$ .
If the photon Planckian spectrum is to be maintained the energy per photon
$\in_\gamma $ must red-shift like the temperature so that $\in_\gamma \sim R^{\bt-1}$.
Then eqs.(5.48) and (5.47) imply that the photon number density\\
 $n_\gamma \propto\rh_\gamma/\in_\gamma \sim R^{3(\bt-1)} $. On the other hand
 the baryon  number density $n_B\sim R^{-3}$. Hence the baryon-to-photon ratio
 \be
 \eta\equiv \fr{n_B}{n_\gamma} \sim R^{-3\bt} \sim T^{3\bt/(1-\bt)} , \ \ \ R\ll R_{eq}.
 \ee
 This is eq.(12) of Freese {\it et al}. .
\se{Cosmic nucleosynthesis}
Primordial synthesis of the light elements nuclei crucially depends on the temperature-time
relation and the expansion rate of the early Universe. In the present model the temperature-time
relation is given by eq.(5.57). The expansion rate, from eq.(5.43), is
\be
 \fr{\dt R}{R}\approx [\fr{\alf^{-1}\rh}{(1-\bt)}]^{1/2},\ \  \ \ \ R\ll R_{eq}
\ee
coinciding with the rate deduced in Ref.[10]. Primordial nucleosynthesis
 depends also on $\eta$ (eq.(5.59) here).
Since eqs.(5.57)-(5.60) coincide with the corresponding relations of  Freese {\it et al}.
$ [10]$ they would lead to these authors nucleosynthesis constraints on $\bt$,
 namely $\bt\le 0.1$ . But the present models require $0 < \bt \le \delta_{rp}$.
 Hence it is consistent with the observed element abundance from primordial
 nucleosynthesis provided $\delta_{rp} < 0.1$ - a condition that readily obtained
 in a present MD universe.
\se{Entropy generation}
 The decreases of $\eta$ with temperature displayed by eq.(5.59) is a consequence of
 the production of the entropy by the vacuum. Requiring that $\eta $ at nucleosynthesis
 falls in the range $10^{-10} \le \eta \le 10^{-9} $, and that at grand unification
 $\eta_{GUT}<10^{-4}$, Freese {\it et al}. $ [10]$ obtain $\bt < 0.12 $, which is
 comparable to the constraint in the preceding section.
\se{Deceleration  parameter}
 Define
 \be
 \xi\equiv \delta_{rp}-\fr{3\bt}{1-4\bt}
 \ee
 By condition (5.38) $\xi > 0 $ . Hence from eq.(5.33) ,
 \be
 0 < q_p < [\fr{1}{2}+\fr{(1-2\bt)(1-4\bt)\xi}{(1-\bt)}][1+\fr{(1-4\bt)\xi}{(1-\bt)}]^{-1} < 1-2\bt .
 \ee
 More precisely since $0<\bt< \delta_{rp}\ll 1$, then $\xi\ll 1 $ and $0< q_p< 1/2 $.
\se{Age of the Universe}
By Equation (5.29) and (5.61)
\be
\fr{1}{(1-2\bt)R_p^2} =\fr{\alf^{-1}\bt_{mp}\xi}{(1-\bt)(1+\om)}\ \ ,
\ee
and from eq.(5.31),
\be
H_{p}^{2}= \fr{\alf^{-1}\rh_{mp}(1+\delta_{rp})}{(1-\bt)}\ \ .
\ee
Let $R=uR_p$ . For all $u\ge u_2=R_2/R_p $ eq. (5.31) can be written as
\be
\dt u^2 =\fr{H_p^2}{(1-\delta_{rp})}[\fr{\xi}{1+\om}+\fr{(1-\bt)}{(1-4\bt)u }+\fr{\om\xi}{(1+\om)u^{2-4\bt}}],
\ee
Taking approximately $t=t_2\approx 0 $ when $u=u_2\approx 0 $ we may write for the
age of the Universe
\be
t_p(\om;  \bt)=\fr{(1+\delta_{rp})^{1/2}}{H_p}\int_0^1u^{1-2\bt}[\fr{\xi u^{2-4\bt}}{1+\om} +\fr{(1-\bt)u^{1-4\bt}}{(1-4\bt)} +\fr{\om\xi}{1+\om}]^{-1/2}du .
\ee
It is readily verifiable that $t_p(\om;\bt), \bt<1/4 $, is a decreasing function
of $\om$ so that
\be
H_pt_p(\om;\bt)< H_pt_p(0;\bt) =(1+b)^{1/2} J,
\ee
where
$$ J = \int_0^1 u^{1/2}[1+bu]^{-1/2} du $$
$$=\fr{(1+b)^{1/2}}{b} +\fr{1}{2b\sqrt{b}}\ln[\fr{\sqrt{1+b}-\sqrt{b}}{\sqrt{1+b}+\sqrt{b}}] $$
and $b=\fr{\xi(1-4\bt)}{(1-\bt)} \ll 1 $. Thus $ J\rightarrow (1+b)^{-1/2}$ or
$ H_pt_p(0;\bt) \rightarrow  1 $ so that one has the upper bound $ t_p(\om;\bt)< H_p^{-1}\approx 10^{42} (\rm GeV)^{-1}
\approx 2\times 10^{10}$yr
\se{Classical cosmological tests}
Returning to eq.(5.31) using eqs.(5.63)  to replace $\rh_{mp}$ by $\Delta $ and
introducing the red-shift $z(1+z=R_p/R)$  we have
\be
\dt R=(1-2\bt)^{-1/2}[a(1+z)[1+\Delta^{-1}(1-\bt)^{-1}(1-4\bt)(1+z)^{1-4\bt}]+1]^{1/2},
\ee
where by eq.(5.63), (5.64) and (5.61),
$$
a\equiv \om(1-\bt)(1-4\bt)^{-1}\Delta=\fr{(1-\bt)(1-2\bt)R_p^2 H_p^2}{(1-4\bt)(1+\delta_{rp} )}
$$
\be
=\fr{(1-2\bt)R_p^2 H_p^2}{(1+b)}
\ee
Now $ \Delta^{-1}=\fr{\om\xi}{1+\om} < \xi\ll 1 $, so that for optical  and radio
cosmic source with $ z\le 5$ ( $z\le 1 $ if normal galaxies and not quasars are considered)
\ [15] ), $ \Delta^{-1}(1-\bt)^{-1}(1-4\bt)(1+z)^{1-4\bt}\ll 1.$
Also $ a\equiv(1+\om)b^{-1}\gg 1 $ , i.e. $ R_p^2 H_p^2 \gg 1 $. This means that although
$\bt \ll 1 $, the $\bt$-term in $\Lambda_p$ is still important, vis-$\grave{a}$-vis the
$\gamma $-term. Thus eq.(5.68) may be approximated simply by
\be
\dt R\approx \sqrt{a}(1-2\bt)^{-1/2}(1+z)^{1/2}.
\ee
The RW metric can be written as
\be
ds^2=d t^2-R^2[d\chi^2+r^2(\chi)[d\theta^2+\sin^2\theta d\phi^2]]
\ee
where $r(\chi)=\sin\chi$ when $k=1$. For light propagation along a radial ( $d\theta=d\phi=0)$ null
geodesic from a distant cosmic source  at $(\chi, \theta,\phi)$ to us at $\chi=0$ eqs.(5.70) and (5.71) give
\be
d\chi = -\fr{dR}{\dt R R}  =\fr{(1-2\bt)^{1/2}dz}{\sqrt{a}(1+z)^{3/2}},
\ee
which upon integration yields ($\chi(z=0)=0$):
\be
\chi =\fr{2(1-2\bt)^{1/2}}{\sqrt{a}}[1-(1+z)^{-1/2}].
\ee
Because $ a \gg 1, \sin\chi\approx \chi . $ Hence
$$\xi \equiv R_p r(\chi)\approx \fr{2(1+b)^{1/2}}{H_p}[1-(1+z)^{-1/2}]$$
\be
 =(1+b)^{1/2}\xi_{ES},
 \ee
 where $ \xi_{ES}$ denotes the product \ $ R_p r(\chi)$ in the
 $ k=0 $ Einstein-de Sitter model.
 This result is important for the following cosmological tests:
\subsection{Magnitude versus red-shift relation }
 The apparent and absolute bolometric magnitudes $m$ and $M$ of
 a galaxy of red-shift  $z$ are related by [16]
 \be
 m - M \equiv \mu =5 log_{10} D +25,
 \ee
 where $\mu$ is the distance modulus and $D=\xi(1+z) $ is the
luminosity distance of the galaxy, measured in $Mpc$.  From eq.(5.74),
 \be
 \mu \approx \mu_{ES}+ \fr{5}{2}\log(1+b).
 \ee
\subsection{Galactic diameters}
 A galaxy of linear dimension $dl$ has angular diameter
 \be
 d\theta=(1+z)\xi^{-1}dl
 \ee
Hence by eq.(5.74),
\be
 d\theta\approx d\theta_{ES} (1+b)^{-1/2}.
 \ee
\subsection{Number counts of sources}
The number of uniformly distributed optical
and radio sources in the volume element $dV$ ,
\be
dV =R^{3} r^{2} (\chi)d\Omega d\chi, \ \ \ \ \  d\Omega=\sin \theta d\theta d\phi ,
\ee
is  \ \ [17]:
\be
dN=n R^3 r^2(\chi)d\Omega d\chi=n_pR_p^3r^2(\chi)d\Omega d\chi
\ee
where $  n(t)\propto \rh_{m}(t)$ is the number density
and $ n R^3\propto \rh_{m}R^3=\rh_{mp} R_p^3 $ so that $ nR^3=n_p R_p^3 $ .
Noting that
\be
\fr{n_p}{n_p^{ES}}=\fr{\rh_{mp}}{\rh_{m}p^{ES}}=\fr{(1-\bt)}{(1+\delta_{rp})}=\fr{(1-4\bt)}{(1+b)},
\ee
and using eqs.(5.80), (5.74), (5.72) and (5.69) we obtain
\be
dN\approx \fr{(1-4\bt)n_p^{ES} \xi_{ES}^2 (1+b)^{1/2} dz d\Omega}{H_p(1+z)^{3/2}}.
\ee
On the other hand in the MD universe of the Einstein-de Sitter model
\be
dN_{ES} =n_p^{ES} \zeta_{ES}^2 R_p^{ES}  d\Omega d\chi=\fr{n_p^{ES}\zeta_{ES}^2 dz d\Omega}{H_p (1+z)^{\fr{3}{2}}}\ \ .
\ee
 Hence
 \be
 dN\approx (1-4\bt)(1+b)^{1/2} dN_{ES}
 \ee
 Equation (5.76), (5.78) and (5.84) are the model's predictions for
 the classical cosmological test. Because, $b$,  \ $\bt\ll 1 $ these results
  are approximately the same as in the ES model.
 \se{Nonsingular model: case 2.}
 Here we will take the case for which $\gamma=1-\bt$ and  $k=1$, so
 that $\rh_0\neq 0$. Equations (5.9), (5.10) and (5.11) become
 \be
 \dt R^2=1-(\fr{R_0}{R})^{2-4\bt}                     \ ,
 \ee
 \be
 \rh=\fr{\alf}{R^2}[1-(1-\bt)(\fr{R_0}{R})^{2-4\bt}]     \ ,
 \ee
 \be
 \rh_v=\fr{\alf}{R^2}[1-\bt(\fr{(R_0}{R})]                  \ ,
 \ee
$ A_0=R_0^{-2+4\bt}$, \ \ \ $\fr{\alf\bt}{R_0^2} > 0 $.\\
And for physical reason we will take $\bt > 0$.
 The rate of change of entropy at temperature  $T$ is given by
 \be
 T\fr{dS}{dR}=-R^3\fr{d\rh_v}{dR}
 \ee
 \be
 T\fr{dS}{dR}=2\alf[1-2\bt(1-\bt)(\fr{R_0}{R})^{2-4\bt}]
 \ee
 The equation shows that  $\fr{dS}{dR} > 0$ for all $R \ge R_0$ and
 thereby solving the entropy problem of the standard model .\\
 The density parameter $\Omega$ defined by\\
$ \Omega =\rh/\rh_c$ , $\rh_c=\alf H^2$ becomes
\be
\Omega=1-\bt+\fr{\bt}{\dt R^2}
\ee
The maximum of $\rh$ is attained at
\be
R=R_{mx}=[2(1-\bt)^{1/2(1-2\bt)}]
\ee
\be
\rh_{mx}=\rh(R_{mx})=\fr{1-2\bt}{2(1-\bt)}\fr{\alf}{R_{mx}^2}
\ee
Using eq.(5.9) one can estimate the value of $R_0$ since
$\dd R=\fr{1-2\bt}{R_0} < R_0^{-1} $ and that signal a limit for a cosmic
 acceleration in the early Universe. Such as limit is of the order of Planck
  mass ($M_{pl}=G^{-1/2}$).
  This Universe accelerates less rapidly that the one with $\gamma=1$ .
  Thus starting with some non zero initial density allows the Universe to be less
  accelerating than with zero one; and with a less maximum density than the
  one with $\gamma=1$.
\se{Radiation and matter}
  In the wake of pure radiation era, the rest mass is generated during the
   period, say, $R_1\le R \le R_2$ $[6]$. For $R > R_2$, after creation of
   the rest mass, the matter energy ($E_m$), $E_m= \rh_m R^3$ stayed constant
   so that $E_m=\rh_m R^3=E_{mp}=\rh_{mp}R^3_p $, where ``p" denotes present-day
   quantities. Thus for the present case with $\Lambda$ given by eq.(5.1) and
   $\gamma=1-\bt, \ \ k=1$ one gets
   \be
   \rh_r= \fr{3\bt E_{mp}}{1-4\bt}R^{-3} +\alf R^{-2}[1+\om(\fr{R_p}{R})^{2-4\bt}]\ ,
   \ee
   where $\om+1= \alf^{-1}\rh_{mp}R_p^2(\delta_p-\fr{3\bt}{1-4\bt})$ and
    $\delta_p=\rh_{rp}/\rh_{mp} $.\\ The radiation energy $E_r$, $E_r=\rh_r R^3$
    is given by
    \be
    E_r=\fr{3\bt E_{mp}}{1-4\bt}+\alf R[1+\om(\fr{R_p}{R})^{2-4\bt}]\ ,
    \ee
\be
\rh_v= \fr{\bt E_{mp}}{1-4\bt}R^{-3}+\alf R^{-2}[1+\fr{\bt\om}{1-\bt}(\fr{R_p}{R})^{2-4\bt}]\ ,
\ee
and
\be
\dt R^2=\fr{\alf^{-1}}{1-\bt }R^2=\fr{\alf^{-1} E_mp}{1-4\bt}R^{-1}+\fr{\om}{1-\bt}(\fr{R_p}{R})^{2-4\bt}\ .
\ee
The acceleration parameter, $q$,
\be
q=\fr{-R\dd R}{\dt R^2}
\ee
The present value is
\be
q_p = \fr{1/2+[\om\alf(1-4\bt)/(1-\bt)]\rh_{mp} R_p^2}{1+[(1-\bt+\om)(1-4\bt)/(1-\bt)]\rh_{mp}R_p^2}\approx 1/2 \ .
\ee
Denote by  $R_{eq}$ the value of R at $t=t_{eq}$, the time when radiation and
matter were balanced in equilibrium
$$ E_r(R_{eq})=E_{mp} .$$
The condition that $E_r$ was  decreasing as $R$ approached $R_{eq}$ would then
imply that
\be
(1-4\bt)\om > (\fr{R_p}{R})^{2-4\bt}\ .
\ee
Thus either  $\om >0 $ and $\bt <1/4 $ as $\om <0 $ and $1/4 < \bt <1/2 $.\\
However, the later case implies that $\dt R^2 <0$, $\rh < 0$ for $R_2 < R < R_{eq}$.
In Ref.[3] one approximately has $\rh_r R^4 $ constant or OT constant for $R < R_{eq}$
 in the radiation and matter universe. In the present case for $ R \ll R_{eq}$ the
 first two terms in eq.(5.95) contribute negligibly to $\rh_r$ and therefore
 \be
 \rh_r\approx \fr{\alf\om}{R^2}(\fr{R_p}{R})^{2-4\bt}\ ,
 \ee
 \be
 \rh_v=\fr{\bt}{1-\bt}\fr{\alf\om}{R^2}(\fr{R_p}{R})^{2-4\bt}\ ,
 \ee
 and
 \be
 \dt R^2=\fr{\om}{1-\bt}(\fr{R_p}{R})^{2-4\bt}\ .
 \ee
 This leads to
 \be
 \rh_v=\fr{\bt}{1-\bt}\rh_r  , \ \ \ \ R\ll R_{eq}\ .
\ee
This is the same relation as postulated by Freese {\it et al}. $[10]$ which suggests that
in the early phase of radiation and matter  $\rh_r$ and $\rh_v$ red-shift at
the same rate. The solution of eqs.(5.102) and (5.103) is
\be
R\sim t^{1/2(1-\bt)}\ ,
\ee
and
\be
\rh_v=\fr{\bt}{1-\bt}\rh_r \approx\fr{\alf\bt}{4(1-\bt)^2 \ \ t^2} \ .
\ee
valid in the early radiation and matter (,i.e. $ R\ll R_{eq} $) epochs.\\
This behaviour of $\rh_v\sim t^{-2}$ has been noted by many workers. Though
the model of $\gamma =1, k=1 $ and $\gamma=1-\bt$ has different features at the early
Universe, they give the same results in the wake of early radiation and matter
phases of the Universe that are identical to Freese {\it et al.}
\se{Phase transition}
We consider, finally, the matter generation period $R_1\le R\le R_2$. For $R \ge R_2$
eq.(5.95) implies $\dd R> 0$ if $\om >0 $ and $\bt <1/4  $ in contrast with $ \dd R > 0$
for $ R< R_1 $ as previously noted. Thus  the appearance of rest-mass ushers in
decelerated expansion of the Universe.\\
Equations (5.25) and (5.26) with $\gamma =1-\bt , \ \ k=1$
\be
\fr{dE}{dR}=2\alf(1-2\bt)+3\bt R^2\rh-3(1-\bt)p R^2
\ee
and using eq.(5.94) one gets
\be
3\int_{R_{0}}^{R_2} R^2[(1-\bt)p-\bt\rh]dR=2\alf(1-2\bt)(R_2-R_0)+E_0-E_2\ ,
\ee
$(E_2=E_m(R_p)+E_r(R_2))$
\be
=2\alf(1-2\bt)(R_2-R_0) +\bt\alf R_0-\rh_{mp}R_p^3-\fr{3\bt}{1-4\bt}\rh_{mp}R_p^3-\alf R_2[1+\om(\fr{R_p}{R})^{2-4\bt}]\ ,
\ee
\be
= - \alf(2-5\bt)R_0-\fr{(1-\bt)}{1-4\bt}\rh_{mp}R_p^3 +\alf(1-4\bt)F(\om)\ ,
\ee
where $$ F(\om)=1-\fr{\om}{1-4\bt}(\fr{R_p}{R})^{2-4\bt} < \ \ 0 \ , $$
as long as $\bt < 1/4 $ the above integral is negative, implying that $3p/\rh \ll 1$
for some values of $R$ in the interval $(R_1, R_2$).
During this period the Universe undergone a phase transition, an era separating the
pure radiation and the radiation and matter epochs. It was noted by Freese {\it et al}.
that the constraint that an early radiation epoch be followed by a matter dominated
era  requires that  $x =\bt < \fr{1}{4}$ in both matter and radiation epochs.
Nucleosynthesis in this model proceeds the same as in the previous one.
\se{Discussion and  concluding remarks}
 Our aim has been to discuss a nonsingular Robertson-Walker cosmological
 model with a varying cosmological constant
$ \Lambda=3\gamma R^{-2} +3\bt H^2 $. This form of $\Lambda$ is due to Carvalho {\it et al}. \ [7].
In the considered model the Universe was initially (t=0) empty and had a minimum scale
factor $ R_0 > 0 $. Subsequently it evolved through consecutive phase of
 pure radiation, matter generation and radiation and  matter.
The absence of the initial singularity requires
$ 2\bt < 1$ and $\fr{1}{2}< \gamma \leq k=1 $. In the very early pure radiation
era $\dd R > 0 $. For  $ R \gg R_0 $ in this era $R\sim t$, compared to
$ R \sim t^{\fr{1}{2}}$ in the standard model.
On the other hand generation of entropy throughout the pure radiation era
requires [see eq.(16)] $\fr{ 2\bt (2\gamma-1)(1-\bt) }{(\gamma-\bt)} < 1 . $
To satisfy this condition we have chosen, for simplicity, $\gamma =1$ and have,
 henceforth, confined our selves to this case .
The model is thus an extension of the \"{O}zer-Taha model($\bt=0, \gamma =1$)
albeit in a different direction than that followed by Ref. \ [6].
General physical consideration and an increasing entropy in the radiation
and matter period place a stronger restriction on $\bt$, viz. $0 < \bt \le \delta_{rp} $,
where $\delta_{rp}$ is the present ratio of radiation to matter energy densities.
This constraint is quite stringent if one assumes, as we have done
here, that the present Universe is matter -dominated so that $\delta_{rp} \ll 1 $.
Under these conditions the pure radiation and the radiation
matter eras are separated by a phase transition period that reverses the sign of
 $\dd R $ from $\dd R \ge 0 $ to $\dd R < 0 $, ushering in decelerated
 expansion. But the precise phase transition equation of state is not known.
In the early radiation and matter universe the model is virtually
indistinguishable from the flat-space decaying-vacuum singular
cosmology of Freese {\it et al}.\ [10]. The condition $ 0 < \bt  <
\delta_{rp}$ means that the nucleosynthesis constraint of Freese
{\it et al}., $\bt \le  0.1 $, is easily satisfied for
$\delta_{rp}\ll 1$. This provides an interesting connection
between the present and early Universe. The  \"{O}zer-Taha model
and the model of Freese {\it et al}. are very dissimilar, as
already noted by the latter authors \ [10]. It is therefore
remarkable that an extension of the one converges to the other.
Freese {\it et al}. make the basic postulate that the vacuum and
radiation energy densities red-shift at the same rate, for large
$R$. This feature emerges here in the very early radiation and
matter era, as a mere by-product of the  approach. In the
cosmology of Carvalho {\it et al}. the Freese {\it et al}.
scenario corresponds to $\gamma=k=0, \bt\neq 0$. Here also $
\bt\neq 0 $, but $\gamma=k=1$. As to the age of the Universe  and
the deceleration parameter, our model predicts $t_p< H_p^{-1} $
and $ 0 < q_p < 1/2 $. On the other  hand, in the singular
cosmology of Carvalho {\it et al}. $\bt$ is a free parameter that
can be adjusted to produce $t_p > H_p^{-1} $ and $ q_p < 0$. We
have also examined the implications of the model for the classical
cosmological tests. The results approach Einstein-de Sitter model
predictions. In the present work the density parameter $\Omega
=1-\bt$ = constant, before and after the phase transition. Hence $
1-\delta_{rp} < \Omega < 1 $ although this implies the absence of
flatness  fine-tuning problem it would apparently require the
existence of dark matter. Finally, the proposed cosmology predicts
a continuously expanding Universe.\\ \se{ References} $[1]$
S.Weinberg, {\it Rev. Mod. Phys.61},1(1989); S.M. Caroll,
W.A.Press, and E.L.Turner, {\it Annu. Rev. Astron. Astrophys. 30},
499 (1992)\\ $[2]$ J.R.Grott, J.E.Gunn, D.N.Schramn, and
B.M.Tinsley, {\it Astrophys. J. 194}, 54 (1974)\\ $[3]$ M.\"{O}zer
and M.O.Taha, {\it Phy.Rev.Lett.B171}, (1986), {\it Nucl. Phys.
B287}, 776 (1987)\\ $[4]$ W.Chen and Y-S.Wu, {\it Phys. Rev.D41},
659 (1990)\\ $[5]$ A-M.M.Abdel Rahman, {\it Gen Rel. Gravit.22},
655 (1990)\\ $[6]$ A-M.M.Abdel Rahman, {\it Phy. Rev.D45},3497
(1990)\\ $[7]$ J.C. Carvalho, J.A.S.Lima and I.Waga, {\it Phy.
Rev.D46}, 2404 (1992)\\ $[8]$ M.Gasperini, {\it Astrophys. Space
Sci.138,} 387 (1987); E.Caianillo, {\it Nouvo Cim. Lett.32}, 65
(1981), G. Marmo and G. Vilasi, {\it Nouvo. Cim. Lett.34}, 112
(1982)\\ $[9]$ M.S. Berman {\it Phy. Rev.D43}, 1075 (1991)\\
$[10]$ K.Freese, F.C. Adams, J.A.Frieman and E.Mottola, {\it Nucl.
Phys. B287}, 797 (1987)\\ $[11]$ A.D.Doglov, in the very early
universe (G.W. Gibson, S.W. Hawking and S.T.C. Siklos (Cambridge
university Press , U.K. 1983, p.449), L.Abbot, {\it Phy. Rev.
Lett.150B}, 427(1985), T.Banks, {\it Phy. Rev. Lett.52}, 1461
(1984), {\it Nucl. Phys. B249}, 332 (1985), see also Berman and
references therein.\\ $[12]$ P.Meszarios, {\it Astron. and
Astrophys. 37}, 225 (1974)\\ $[13]$ This assumption is quite
plausible in the very early radiation and matter universe because
, with $\beta\ll 1$, is approximately the Euclidean Friedmann
equations\\ $[14]$ M.Rowan-Robinson, Cosmology (Oxford University
Press, U.K 1981, p.49)\\ $[15]$ J.V.Narlikar, Introduction to
Cosmology (Cambridge University Press, U.K., 1993)
    pp.291-293\\
$[16]$ See Rowan-Robinson [15], p.109\\
$[17]$ A.S.Myskis, Advanced Mathematics for Engineers, English translation by
V.M.Volsov and I.G.Volsova (Mir Publishers, 1975, pp.638-639)\\
$[18]$ M.V.Berry, Principle of Cosmology and Gravitation (Adams Hilger, New York, 1989, p.146)
\chapter{A Nonsingular Viscous Cosmological Model}
\se{Introduction}
In one approach for solving the cosmological constant problem [1, 2] cosmologies
with decaying vacuum energy  are introduced ( $[3, 4]$ and references therein).
But in most of the proposed models dissipative effects that were probably very important
in the early Universe are not considered.
On the other hand the influence of bulk viscosity on cosmic evolution in standard Friedmann
models has been discussed in many works( for an excellent review see $[5]$). In these
works the coefficient of bulk viscosity  $\eta$, say, is usually taken to have a
power law dependence on the cosmic energy density $\rho$, viz. $\eta \propto \rho^n $,
$n$ constant.
Recently Calvao, de Oliveira, Pav$\acute{o}$n and Salim $[6]$ generalized the model of Chen and Wu
$[7]$, in which the vacuum energy decreases with cosmic expansion as the inverse square
of the Robertson-Walker (RW) scale factor, by appending to the energy-momentum
tensor a bulk viscosity term with $\eta \propto \rho$. The resulting field equations in
this case are found to have no analytical solutions in general and methods of qualitative
analysis of differential equations are used to deduce the cosmological implications
 of the model.\\
The present paper also attempts to incorporate bulk viscosity in a decaying vacuum cosmology.
But unlike Ref. $[6]$ we take $n =1/2$ in the dependence of $\eta$ on $\rho$, i.e. $\eta \propto \rho^{1/2}$.
For this case the solutions of the field equations for flat cosmological models with
 vanishing cosmological constant $\Lambda$ appears to have a special status:\\
 They are the only solutions with structural stability $[8]$ . In addition, Beesham
$ [9]$ has shown that the $n=1/2$  models are equivalent to the flat variable-$\Lambda$
 cosmologies of Berman $[10]$. Because of these peculiarities of the models with
 $\eta \propto \rho^{1/2} $ we limit our self to this case.\\
 We also assume a Universe  with exactly the critical  density. Consequently
 the vacuum energy decays as in the \"{O}zer - Taha model $[11]$. The critical density assumption,
which is indicated by inflation, is theoretically appealing because the apparent
closeness of the present and critical densities would else be difficult to
understand on other than anthropic grounds.

Our model which is nonsingular and closed  is introduced in Sec.5.2 . Several implications
 of this model turn out to be equivalent to the corresponding  consequences in
 previously proposed  non viscous cosmologies  $[12-14]$.
 \se{The Model}
    In a homogeneous isotropic universe with a perfect fluid energy-momentum tensor,
Einstein's  equations with a variable cosmological `constant' $\Lambda$
 give ($\kappa\equiv 3/8\pi G$)[11]:
\be
\fr{\dot R^2}{R^2}=\kappa^{-1}\rho +\fr{\Lambda}{3}-\fr{k}{R^2},
\ee
\be
\fr{d(\rho R^3)}{dR}+3pR^2+\fr{\kappa}{3}R^3\fr{d\Lambda}{dR}=0 ,
 \ee
where $\rho$  and  p are the cosmic energy density and pressure, respectively, and k the curvature
index.
To obtain the field equations with bulk viscosity we replace $p$ by the effective pressure [15]
\be
p^* = p-3\eta H,
\ee
where $H=\fr{\dot R}{R}$ is the Hubble's constant and $\eta $ is the coefficient
 of bulk viscosity.\\
 As remarked in the introduction we set $\rho $ equal to the critical density.
\be
\rho=\rho_c=\kappa H^2,
\ee
throughout cosmic evolution. Consequently eq.(6.10) gives the \"{O}zer-Taha $[11]$
result
\be
 \Lambda =\fr{3k}{R^2}
\ee
 The form  $\Lambda \propto R^{-2} $ has also been suggested by Chen and Wu $[7]$ on the basis
 of dimensional arguments consistent with quantum gravity. Pav$\acute{o}$n $[16]$, on the other
 hand, has used the Landau-Lifshitz fluctuation theory to study the physical
 consistency of several  $\Lambda $ variations and has concluded that  $\Lambda$
 should vary as  $ R^{-2}$. The detailed observational consequences  of this behaviour
 have been examined in Refs. $[7, 11, 13 , 17, 18]$.
 As mentioned in the introduction we choose $\eta \propto \rho ^{1/2} $.
 Specifically we write
\be
\eta = \fr{\eta_0}{2(6\pi)^{1/2}}\rho^{1/2},   \ \ \ \ \ \eta_0 \ge 0 \ \ \ ,\rm const..
\ee
Together with eq.(6.13) and the usual equation of state\\
 $ p=(\gamma-1)\rho $ ,
$0\le\gamma\le 2$ this leads to
\be
p^* = [\gamma-1-\fr{\eta_0}{M_{Pl}}]\rho,
\ee
where  $M_{Pl}= G^{-1/2}$ is the Planck mass.
Equation (6.11), with $p \rightarrow p^*$, and eqs.(6.16) and (6.14) give
\be
\fr{d\rho}{dR}+\fr{3}{R}[\gamma-\fr{\eta_0}{M_{Pl}}]\rho=\fr{2\kappa k}{R^3},
\ee
which has the solution
\be
\kappa^{-1}\rho R^2=\dot R^2=\fr{2k}{3\gamma-2-4\beta}+AR^{-3\gamma+2+4\beta}  ,
\ee
where $\beta=\fr{3\eta_0}{4M_{Pl}}$ and $A$ is a constant. In particular for the radiation-
($\gamma=4/3)$ dominated (RD) phase of the Universe
\be
\dot R^2=\fr{k}{1-2\beta}+AR^{-2+4\beta}.
\ee
Carvalho, Lima and Waga $[12]$ have proposed a model Universe based on the cosmological
constant ansatz
\be
\Lambda=3\beta H^2+\fr{3\alpha}{R^2}  \  ,
\ee
where $\alpha$ and $\beta$ are dimensionless  constants of order unity. The detailed cosmological
implications of this postulate have been explored in $[12, 3, 14]$. In particular
a nonsingular scenario based on eq.(6.19) was studied in $[14]$.\\
It follows from eq.(6.20) that in the RD universe $[12]$
\be
\dot R^2 = \fr{2\alpha-k}{1-2\beta}+A_0R^{-2+4\beta},
\ee
and
\be
\rho=\fr{\kappa(\alpha-\beta k)}{1-2\beta}R^{-2} +\kappa(1-\beta)A_0 R^{-4+4\beta}.
\ee
For $\alpha=k$ eqs.(6.21) and (6.19) are formally equivalent. It is therefore interesting
to ask to what extent their equivalence is reflected  in the corresponding cosmologies.
To answer this question we consider a nonsingular scenario as in Ref. $[14]$.\\
A Universe with a non vanishing minimum scale factor $R_0$ at $t=0$ arises from eq.(6.19)
 if $ A < 0 $, $\beta < 1/2 $ and $k=1$ (closed universe). Then
\be
 \rho=\fr{\kappa\dot R^2}{R^2}=\fr{\kappa}{R^2(1-2\beta)}[1-\fr{R_0^{2-4\beta}}{R^{2-4\beta}}],
 \ee
 with $\rho_0 = 0 $ and $\rho $ maximum $(=\fr{\kappa}{2(1-\beta)R^2_{mx}}) $ at
$$ R=R_{mx}=[2(1-\beta)]^{\fr{1}{2-4\beta}}R_0.$$
 An estimate of $R_0$ may be obtained as follows. From eq.(6.19) $ 0 <\ddot R \le R_0^{-1}$
 so that $ R_0^{-1}$ represents a cosmic acceleration limit.  Such a limit of the order
 of $M_{Pl}$ has been suggested before $[19]$. Taking $R^{-1}_0 \sim M_{Pl} $ yields
 $ R_0 \sim 10^{-33} \rm cm .$     \\
 We next investigate the implications of the model in the cosmic phase of noninteracting
 radiation of density $\rho_r$ and pressureless non relativistic matter of density $\rho_m$
$ (\rho_r+\rho_m=\rho)$.\\  We assume that the vacuum couples to radiation only $[11, 13, 14, 18]$.
Hence $p=\rho_r/3$ and the matter energy $ E_m = \rho_m R^3 = \rho_{mp}R^3_p=E_{mp} $,
where subscript ``p'' denotes present-day quantities. Then eqs.(6.11), (6.14) and (6.16) give (k=1)
\be
\fr{d\rho_r}{dR} + \fr{4(1-\beta)}{R}\rho_r = \fr{2\kappa}{R^3} + \fr{4\beta}{R^4}E_{mp}
\ee
which has the solution
\be
\rho_r= \fr{4\beta E_{mp}}{(1-4\beta)R^3} + \fr{\kappa}{(1-2\beta)R^2}\left [ 1+\fr{\om R_{p}^{2-4\beta}}{R^{2-4\beta}}\right ],
\ee
where
\be
(1+\om)= (1-2\beta)\kappa^{-1}\rho_{mp}R_{p}^2[\delta_p-\fr{4\beta}{1-4\beta}],
\ee
$\delta_p=\fr{\rho_{rp}}{\rho_{mp}}$ being the present ratio of radiation-to-matter energy
densities. It follows from eq.(6.24) and conservation of non relativistic matter that
\be
\dot R^2= \kappa^{-1}\rho R^2=\fr{\kappa^{-1}\rho_{mp}R_p^3}{(1-4\beta)R}+\fr{\om R_p^{2-4\beta}}{(1-2\beta)R^{2-4\beta}}+(1-2\beta)^{-1}.
\ee
Denote R at $t=t_{eq}$, the time when radiation and matter were equal, by $R_{eq}$,
i.e. $E_r(R_{eq})=E_m(R_{eq})=E_{mp}$, where $ E_r(R)=\rho_r R^3$ is the radiation energy. Then the condition that  $E_r$ was decreasing
 as R approached  $R_{eq} $ leads to
\be
(1-4\beta) \om > \fr{R_{eq}^{2-4\beta}}{R_{p}^{2-4\beta}}.
\ee
Implying that either
$\om > 0  \  \ and \ \   \beta < 1/4 $ or $ \om < 0 $ and $\beta > 1/4$. But
the latter case would, by virtue of eq.(6.26), imply that $ \dot R^2 < 0$ for $R\le R_{eq}$
and is therefore excluded .\\
Equation (6.27) is formally identical with eq.(6.24) in Ref. $[14]$. But whereas $\beta $
there determines the coefficient of $ H^2$ term in the cosmological constant  ansatz,
here it is a measure of bulk viscosity .\\
 The condition $\beta < 1/4$ implies, from eq.(6.26), that $ \ddot R < 0 $ compared to
 $ \ddot R >0 $ in the pure radiation era. Assuming that a period of rest-mass generation  $R_1\le R\le R_2 $ , say, separated the
  pure radiation  and the matter  and radiation epochs one can readily
  show, in exactly the same way as in $[14]$, that this period is  a phase transition
  era  during which the pressure  becomes small or negative.\\
 The conditions  $\om > 0, \ \ \beta < 1/4 $ and eq.(6.25) lead to $ 4\beta < \delta_p $.
 Hence if the present Universe is matter-dominated  (MD) $\fr{\eta_0}{M_{Pl}}< 4\delta_p/3\ll 1$ , or, from eq.(6.16), that the
 present bulk viscosity contribution to the effective pressure $p^*$ is at most of the order
 of the radiation pressure.
 To examine nucleosynthesis in the present  cosmology  we proceed as follows. Noting that
 $\rho_{mp}R_p^3=\rho_{r,eq}R^3_{eq} $ where $ \rho_{r,eq}$ is the radiation density
 at equilibrium we obtain from eqs.(6.26), (6.24) and (6.27) the approximate relation
\be
\dot R^2 \approx \fr{\om R_p^{2-4\beta}}{(1-2\beta)R^{2-4\beta}}\approx \kappa^{-1}\rho_r R^2,
\ee
valid for $R\ll R_{eq}, \beta \ll 1 $ and $\rho_m\ll \rho_r $ .\\
Equation (6.28) is approximately the standard ($k=0$) model's cosmic expansion rate.
Counting three neutrino species one therefore has for the  neutron-to-proton
``freeze-out " temperature $T_{F10}(\equiv T_F/10^{10} \rm K)\approx 1.11 $, the same as in
standard cosmology. Also from eq.(6.28) $\dot \rho_r/\rho_r\approx -4(1-\beta)\dot R/R\approx-4\dot R/R $
leading to the standard temperature-time relation.\\\
From eqs.(6.28) and (6.25), using $\rho_r = \fr{\pi^2}{30}g_{eff} T^4 $, one has
$$R^{-1}R_p=[\fr{1+\om}{\om\Delta}]^{1/4(1-\beta)} [\fr{T}{T_p}]^{1/(1-\beta)}$$
\be
\approx [\fr{1+\om}{\om\Delta}]^{1/4}\fr{T}{T_p},
\ee
where
\be
 \Delta \approx 1-4\beta\delta_p^{-1}.
 \ee
Denoting the baryon number density by $n_B$ we then have
\be
n_B \approx n_{Bp}[\fr{1+\om}{\om\Delta}]^{3/4} [\fr{T}{T_p}]^3
\ee
for $T\le T_N $, the nucleosynthesis temperature, and where
$n_{Bp}=\kappa m_N^{-1}\Omega_B H_p^2$, with  $m_N$ being the nucleon mass and $ \Omega_B$
the present baryonic fraction of the critical density. The equation determining
$ T_N $(in$\rm  GeV$ ) therefore becomes
\be
\fr{B_d}{T_N}+\fr{3}{2}\ln T_N +\ln(\bar\Omega_B h^2)-14.47 =0,
\ee
 where $B_d(=2.23 \rm MeV )$ is the deuteron binding energy, $h$ is the normalized Hubble
 constant ($0.4 \le h\le 0.8 \ \  [20]) $ and
\be
 \bar\Omega_B=[\fr{1+\om}{\om\Delta}]^{3/4}\ \ \Omega_B\ge\Omega_B,
 \ee
 with the equality sign holding for the non viscous ($\Delta \rightarrow 1$)
  standard ($\om \rightarrow \infty $) case.\\
 Apart from the replacement $\Omega_B\rightarrow \bar\Omega_B $, eq.(6.32) coincides with
the  corresponding standard model relation  $[21]$. Requiring $\bar\Omega_B $ to
satisfy the standard model constraints on $\Omega_B$, i.e. $[22]$
\be
 0.015\le\bar\Omega_B\le0.070
 \ee
 gives $T_N$(standard). With $T_F$ and $T_N$ as in standard cosmology and the nuclear
 physics aspects of nucleosynthesis unaltered the standard nucleosynthesis scenario
 is reproduced.
 The condition (6.34) implies that
\be
\fr{\Omega_B}{0.070} \le \Delta^{3/4}\le 1,
\ee
irrespective of the size of the contribution of vacuum energy. In the presence of
viscosity ($\Delta < 1$) the constraint (6.35) implies a stronger upper  bound on
the baryonic density than required by the standard model.
\se{Concluding remarks}
We have presented a homogeneous isotropic cosmological model with decaying vacuum energy and
bulk viscosity. Our aim has been to generalize an earlier work by Calvao {\it et al}.
$[6]$ in which the bulk viscosity coefficient depends  linearly on the
cosmic energy density $\rho$. In the present model it varies as $\rho^{1/2}$ .\\
Our model is nonsingular, closed and, in several aspects, equivalent to previously proposed
non viscous variable-$\Lambda$ models $[12-14]$. This type of equivalence of viscous and
non viscous variable-$\Lambda$ cosmologies has been noted and discussed by Beesham $[9]$.\\
A noteworthy feature of the present work is that primordial nucleosynthesis proceeds
as in standard cosmology provided the Universe today is matter dominated. This condition
implies that the present bulk viscosity contribution to the cosmic pressure is, at
most, of the order of the radiation pressure. Another consequence  of nucleosynthesis is that
the upper limit on the baryon density is lower than the standard value.
This strengthens the case for nonbaryonic dark matter.
\se{References}
$[1]$ S. Weinberg, {\it Rev. Mod. Phys. 61} (1989) 1.\\
$[2]$ S. M. Carroll, W. H. press and E. L Turner, {\it Annu. Rev. Astron. Astrophysics 30 }(1992) 499.\\
$[3]$ I. Waga, {\it Astrophys. J. 414} (1993) 436.\\
$[4]$ J. A. S. Lima and J. M. F. Maia, {\it Mod. Phys. Lett. A, 8} (1993) 591.\\
$[5]$ O. Gron, {\it Astrophys. Space Sci. 173} (1990) 191.\\
$[6]$ M. O. Calvao, H. P. de Oliveira, D. Pav$\acute{o}$n and J. M. Salim, {\it Phys. Rev. D. 45} (1992) 3869.\\
$[7]$ W. Chen and Y. S. Wu, {\it Phys. Rev. D. 41} (1990) 695.\\
$[8]$ Z. Golda , M. Heller and M. Szydlowski, {\it Astrophys. Space Sci. 90} (1983) 313.\\
$[9]$ A. Beesham, {\it Phys. Rev. D. 48} (1993) 3539.\\
$[10]$ M. S. Berman, {\it Phys. Rev. D. 43} (1991) 1075.\\
$[11]$ M. \"{O}zer and M. O. Taha, {\it Phys. Lett. B171} (1986) 363; {\it Nucl. phys. B287} (1987) 776.\\
$[12]$ J. C. Carvalho, J. A. S. Lima and I. Waga, {\it Phys. Rev. D. 46} (1992) 2404.\\
$[13]$ A. -M. M. Abdel-Rahman, {\it Phys. Rev. D. 45} (1992) 3497.\\
$[14]$ A. I. Arbab and A. -M. M. Abdel-Rahman, {\it Phys. Rev. D. 50} (1994) 7725.\\
$[15]$ S. Weinberg, Gravitation and Cosmology (Wiley, New York, 1972).\\
$[16]$ D. Pav$\acute{o}$n,{\it Phys.Rev. D. 43} (1991) 375.\\
$[17]$ A. -M. M. Abdel-Rahman, {\it Gen. Relativ. Gravit. 22} (1990) 655; Gen. Relativ. Gravit.. 27(1995) 573\\
$[18]$ T. M. Abdalla and A. -M. M. Abdel-Rahman, {\it Phys. Rev. D.46} (1992) 5675.\\
$[19]$ M. Gasperini, {\it Astrophys. Space Sci. 138} (1987) 387; L. C. Garcia de Andrade, Nouvo Cimento B. 109 (1994) 757.\\
$[20]$ W. Freedman {\it et al}, Nature 371 (1994) 433.\\
$[21]$ P. J. E. Peebles, Physical Cosmology (Princeton University Press, Princeton, NJ, 1971) pp. 240 - 277.\\
$[22]$ P. J. Kernan and L. M. Krauss, {\it Phys. Rev. Lett. 72} (1994) 3309.
\chapter{A Viscous Universe with Variable $G$ and $\Lambda$}
\se{Introduction}
The role of viscosity in cosmology has been studied by several authors [1-4].
The bulk viscosity associated with grand unified theory phase transition
can lead to inflationary universe  scenario.
It was well known that in an early stage of the Universe when
neutrino decoupling occurred, the matter behaves like viscous fluid \ [22].
The coefficient of viscosity is known to decrease as the Universe expands.
Beesham \ [20] studied a universe consisting of a cosmological constant
($\Lambda \sim t^{-2} $) and bulk viscosity. He showed that the Berman
model could be viscous model for $n=1/2$ .\\
        More recently Abdel Rahman considered a model in which  the
gravitational  constant, $G$, varies with time  but energy was conserved \ [11].
Other models have been considered in the literature by Sistero and Kalligas
{\it et al }. In the present work we will investigate the effect of viscosity in a
universe where $G$ and $\Lambda$ vary in such a way that energy is
conserved .
\se{ The Model}
In a Robertson Walker universe
\be
d\tau^{2}= dt^{2}-
R^{2}(t)[\fr{dr^{2}}{1-kr^2}+r^{2}(d\theta^{2}+\sin\theta^{2}d\phi^{2})],
\ee
where $k$ is the curvature index.\\
Einstein's field equations with time dependent cosmological and gravitational
``constants''
\be
 R_{\mu\nu}-\fr{1}{2}g_{\mu\nu}R=8\pi GT_{\mu\nu}+ \Lambda g_{\mu\nu} ,
\ee
and the perfect fluid energy momentum tensor \\
\be
T_{\mu\nu}=(\rho+p)u_{\mu}u_{\nu}-pg_{\mu\nu} \ ,
\ee
yield the two independent equations
\be
3\fr{\ddot{R}}{R}= -4\pi G(3p+\rho-\fr{\Lambda}{4\pi G})\ ,
\ee
\be
3\fr{\dot{R}^{2}}{R^2}=8\pi G(\rho+\fr{\Lambda}{8\pi G})-\fr{3k}{R^2}  \ .
\ee
Elimination of $\ddot{R}$  gives
\be
3(p+\rho)\dot{R}=-(\fr{\dot{G}}{G}\rho+\dot{\rho}+\fr{\dot{\Lambda}}{8\pi G})R\ .
\ee
The conservation of energy and momentum yields
\be
3(p+\rho)=-R\fr{d\rho}{dR}                                                     .
\ee
The effect of bulk viscosity in the field equation is to replace
$p$ by $p-3\eta H$, where $\eta$ is the viscosity coefficient. It follows
immediately that
\be
9\eta H\dot{R}=(\fr{\dot{G}}{G}\rho+\fr{\dot{\Lambda}}{8\pi G})R,
\ee
and
\be
\dot{\rho}+3H(\rho+p)=0                    \    .
\ee
eq.(8) can be written as
\be
9\eta \fr{H}{R}=\fr{G'}{G}\rho+\fr{\Lambda'}{8\pi G} \ ,
\ee
where prime denotes derivative w.r.t. scale factor $R$ while dot is the
derivative w.r.t to cosmic time $t$. In what follows we will consider a flat universe, $k=0$\\
Equation (7.5) and (7.17) lead to
\be
8\pi G\rho=3(1-\bt)H^2                                 \   ,
\ee
 and the equation of state
\be p=(\gamma-1)\rho \ee in eq.(7.8) and (7.9) lead to
\be
\rho=AR^{-3\gamma}                                      \      ,
\ee
where A is a constant.
\be
9\eta\fr{H}{R}=2\fr{H'}{H}\rho-\rho'+2\beta\fr{H'}{H}\rho\  ,
\ee
or \be
\fr{H'}{H^{2}}+\fr{3\gamma}{2(1+\beta)R}\fr{1}{H}=\fr{9\eta_{0}A^{n-1}}{2(1+\beta)}R^{-3\gamma n+3\gamma-1}\ ,
\ee
where we have taken the viscosity  coefficient to have the power law
\be
\eta=\eta_{0}\rho^{n} , \ \ \ \eta_{0} \ge 0 \ \ , n\ \ \rm \ const.
\ee
and the ansatz [30]
\be
\Lambda =3\beta H^{2},\ \ \ \beta \ \ \ \rm const..
\ee
 The solution of eq.(7.15) is obtained as follows\\
Let $ y=\fr{1}{H}$ and\ \ \ $ a=\fr{3\gamma(1-\bt)}{2}$.
Therefore
\be
\fr{d}{dR}yR^{-a}=\fr{-9(1-\bt)\eta_0A^{n-1}}{2}R^{-3\gamma n+3\gamma-a-1},
\ee
\be
yR^{-a}=\fr{9(1-\bt)\eta_0A^{n-1}}{2(3\gamma n-3\gamma +a)}R^{-3\gamma n+3\gamma -a},
\ee
\be
y= \fr{9(1-\bt)\eta_0A^{n-1}}{2}{(3\gamma n-3\gamma+a)}R^{-3\gamma n+3\gamma}        ,
\ee
and finally
\be
H= \fr{2(3\gamma n-3\gamma +a)}{9(1-\bt)\eta_0A^{n-1}}R^{3\gamma n-3\gamma }          .
\ee
eq.(7.11) and (7.21) give
\be
G=\fr{3D^2(1-\bt)}{8\pi A}R^{3\gamma(2n-1)}                                      ,
\ee
where\\
$$D=\fr{2(2n-1-\bt)}{9(1-\bt)\eta_0A^{n-1}}\ ,$$
eq.(7.21) gives
\be
R(t)=[3D\gamma(1-n)]^{\fr{1}{3\gamma(1-n)}}t^{\fr{1}{3\gamma(1-n)}}\ .
\ee
Hence eq.(7.13) and (7.22) become
\be
\rho(t)=A't^{\fr{-1}{1-n}}                          \     ,
\ee
\be
G(t)=B't^{\fr{2n-1}{1-n}}                            \    ,
\ee
and
\be
\eta(t)=A_0 t^{\fr{-n}{1-n}}                          \   ,
\ee
where\\
$A'=A[3m\gamma(1-n)]^{\fr{-1}{1-n}},$\\
 $B'=\fr{3m^2}{8\pi A}[3m\gamma(1-n)]^{\fr{2n-1}{1-n}}\ ,$ and
$A_0=A'^n\eta_0$.\\
 The Hubble parameter is
 \be
 H(t)=\fr{1}{3\gamma(1-n)}\fr{1}{t}
 \ee
 where \ \ \ $ 0 \le n \le 1$.
 This condition on $n$ rules out some  models with $n > 1 $ \ [3].
 The cosmological constant becomes
 \be
 \Lambda=\fr{\beta}{3\gamma^2(1-n)^2}\fr{1}{t^2}
 \ee
 This law of variation of $\Lambda$ is thought to be fundamental \ [20].
 The vacuum energy density ($\rh_v$) is given by
 \be
 \rh_v=\fr{\Lambda}{8\pi G} \ ,
 \ee
and from eq.(7.11) and (7.17) we obtain
 \be
 \rh_v=\fr{\bt}{1-\bt}\ \rho .
 \ee
For an expanding Universe, i.e. $H > 0$, we must have $D > 0$. This implies that
 $$ 2n-\bt-1 > 0 ,$$ or
\be
\bt < 2n -1.
\ee
Whether $G$ increases or decreases depends on the value of $n$. For $ n > 1/2$  \ $G$ increases
with time and for  $ n < 1/2$ \ $G$ decreases with time and for $n = 1/2$ \ $G$ remains constant.
The condition (7.31) now gives\\
$ \Lambda > 0, \ \ \ \ G $ increases with time\\
and                            \\
$  \Lambda < 0, \ \ \ \ G $ decreases with time or remains constant.\\
To solve the persisting age problem of the ES model we must  have
$t\ H > 2/3 $. This result require that $G$  is an increasing function of time.
Recently, Massa (1995) proposed a model in which $G$ increases with time.
An increasing $G$ would cause the Planck length $l_P=\sqrt{G}$\ \ to be an increasing
function of time, and the quantum fluctuations on the metric would be vanishingly
small in the very early Universe. A fully classical description of the Universe for
all $t> 0$\ would be possible. This is perhaps one of the reasons to consider the
increasing $G$ model [6].

 Following Freese {\it et al}. one can put a stringent constrain on  the value of $\bt$ .
 The parameter $x$ of Freese {\it et al}. is equivalent to  $\bt$
and since $x \le 0.1 $ this implies   $\bt \le 0.1 \ \ $ for the
nucleosynthesis constraints to hold.
 The deceleration parameter is given by\\
$$ q= -\fr{R\ddot{R}}{\dot{R}^2}\ \ ,$$
\be
q=3\gamma -3\gamma n-1               \   .
\ee
This shows that the deceleration parameter is constant.
The constant deceleration models have considered by Berman and Som \ [9,15].
Equation (7.27) can be written as
\be
H= \fr{1}{(1+q)}\fr{1}{t}\ ,
\ee
and for the present phase ``p "
\be
t_p=\fr{1}{(1+q_p)}\fr{1}{H_p} \ .
\ee
It is evident that negative $q_{\rm p}$ would increase  of the present age of the Universe.\\
From eq.(7.25) we obtain
\be
\fr{\dot G}{G}=\fr{2n-1}{1-n} \ \fr{1}{t}\ ,
\ee
and the present value is
\be
 (\fr{\dot G}{G})_p=\fr{2n-1}{1-n}\fr{1}{t_p}=\fr{2n-1}{1-n}(1+q_p)H_p.
\ee
A power law dependence of $G$ was obtained by Kalligas {\it et al.}[19], it has been shown
to lead naturally to $\Lambda \sim t^{-2}$.
Unlike the model of Abdel Rahman and Beesham, this model shows a constant
$G$ does not imply constant $\Lambda$.
We see that the quantity $G\rho$ satisfies the condition for a Machian
cosmological solution, i.e. $G\rho \sim H^2$, (see [25]). This also follows from the model of Kalligas {\it et al }.\\
The relationship between  our model and that due to Kalligas {\it et al.} is manifested in the following replacement\\
$$n=\fr{1+n_{\rm K}}{2+n_{\rm K}}$$
and $$\beta=\fr{n_{\rm K}}{2+n_{\rm K}}$$
where\\
$ n_{\rm K}$ : $n$ due to Kalligas {\it et al }.
This furnishes the resemblance. Hence Kalligas {\it et al.} model is
equivalent to a viscous model .
\se{The horizon problem}
The horizon distance,i.e. the size of the causally connected region, is given by
$$ d_H=R(t)\int_{t_0}^{t}\fr{dt'}{R(t')},$$
$$ d_H(t,t_0)=\fr{3\gamma-3\gamma n }{3\gamma-3\gamma n-1}(t,t_0) .$$
We would like to have $3\gamma -3\gamma n=1$, so
 $$n=\fr{3\gamma-1}{3\gamma}$$\\
Note that $ 3\gamma-3\gamma n > 0 $ implies $ n< 1 $.\\
In what follows we will discuss some classes of models.
\se{Model with n=1}
Equation (7.7) becomes
\be
\fr{d}{dR}yR^{-a}=-\fr{9(1-\bt)\eta_0A^{n-1}}{2}R^{-1-a} \  ,
\ee
\be
yR^{-a}=\fr{9(1-\bt)\eta_0A^{n-1}}{a}R^{-a}               \ ,
\ee
\be
y=\fr{9(1-\bt)\eta_0A^{n-1}}{a}=\rm const                  \  ,
\ee
or
\be
H=\fr{a}{9(1-\bt)\eta_0A^{n-1}}=\fr{\gamma}{3\eta_0}\equiv H_0 \ .
\ee
Hence
$$ R(t)=F\exp H_0 t  \ \ , \ \ F\ \ \rm const., $$
which is an inflationary solution.\\
Such a solution has been obtained by several authors \ [2,5,13].
Here the density is not constant but has the following  variation
$$\rho=AF^{-3\gamma}\exp-3\gamma H_0 t.$$
Such a solution was obtained by Berman and Som for the Brans-Dicke theory for the scalar
field $\phi$ where $\phi= \ 1/G $  \ [26].\\
In the present case, however, $G$ is not constant during this epoch, viz.
$$ G(t)=M\exp 3\gamma H_0 t,$$
where
$$ M=\fr{3H_0^2F^{3\gamma}(1-\bt)}{8\pi A}.$$
\se{Model with $n=1/2$, \ \ $\gamma=1$}
Equations(7.23)--(7.25) become
$$ R(t)=F t^{\fr{2}{3}}, \ \ \  F \ \ \rm  const.$$
$$ \rho=A_0t^{-2} , \ \ \ A_0 \ \ \rm\ const.$$
and
$$ G=\rm const.$$
The Hubble parameter is $ H(t)=\fr{2}{3}t^{-1} $.
This is the flat FRW universe result.\\
The deceleration parameter is
$$ q=-3n+2=1/2 $$
Since several authors claim that the age of the Universe computed from
the FRW flat model tends to be smaller than the range given
by observation, $ 0.6< H_pt_p< 1.4 $, our model  could give a better
value for any departure from $n=1/2$ .\\
However, it was found that only $n=1/2$ solution are structurally stable \ [21].\\
It was shown by Beesham that Berman solution (a power law for $R$) is a viscous solution
with $n=1/2$. The relationship between our model and Berman's [9] is
$$ m=3\gamma(1-n)\ \ .$$
The value of  $m$ in our case is not put by hand, but emerges naturally
from the dependence of the  viscosity on the energy density($ \eta \sim \rho^n $)
in a given epoch.
This solution seems  more elegant.
\se{Model  with n=0 , \ \ $\gamma=1$}
Equations(7.23)--(7.25) give
$$R(t)=F't^{1/3},$$
$$\rho=A't^{-1},$$
and
$$G(t)=Bt^{-1},$$
 where $F'$, $A'$ and B are constants.
  This is a model of constant bulk viscosity. It resembles the Brans-Dicke model
[16]. We see that
\be
\fr{\dot G}{G}=-\fr{1}{t}=-3H^{-1}\ ,
\ee
\be
(\fr{\dot G}{G})_p=-3H^{-1}_p  \    .
\ee
This solution was obtained by Berman [15] for the Bertolami equation for the present
phase.\\
Note that in GR, $k=0 $
\be
\rh=\fr{1}{6\pi Gt^2}  .
\ee
Whether our result is acceptable or not depends upon the value we  measure
for  $\ \ (\fr{\dot \rho}{\rho})_p $ for the present phase.\\
This also resembles the Dirac no creation model. For this class of solutions $q=2$.\\
\se{Model with n=2/3, $\gamma =1$ }
The scale factor is given by
$$R(t)=R_0 t  \ \ \ \ R_0  \ \rm const.,$$
 and
 $$\rho=A' t^{-3},$$
$$ G=Bt                 ,$$
where $A'$ and B are constants.

 This linear variation of $G$ has been found by Berman \ [15] for the Bertolami
solution for the Brans-Dicke theory (BD) with a time varying cosmological constant for
the present phase. For this model $q=0$.
\se{Model with n=3/4, \ \ $\gamma =4/3 $ }
The scale factor is given by
$$ R(t)=F t,$$
and
$$ \rho=A't^{-4},$$
$$ G=Bt^{2}  ,$$
where F, $A'$ and B are constants.\\
This solution was obtained by Berman \ [15] for the Bertolami theory for the
radiation era. He also found that $T \propto R^{-1}$, preserving Stefan's law.
It was also found by Abdel Rahman that a variable $G$ and $\Lambda$ model
lead to a similar result for the radiation universe [11]. In his model he
considered $\Lambda \sim R^{-2}$. For this class of models $q=0$.

More recently, (1995) Massa has considered a model which  support an increasing
$G$ constant. In his work his considered a ``maximal power hypothesis (MPH) in the
Einstein-Cartan theory of gravity. Equivalence of his work and ours requires
$n > 1/2$.
\se{Model with n=1/2 , \ \ $\gamma=4/3$}
For this model
$$ R(t)=F t^{1/2} \ \ ,$$
$$ \rho= A't^{-2} \ \ ,$$
and
$$ G= B = \rm const., $$
where F, $A'$ and B are constants.\\
This special value for $ n$ gives a constant G in both radiation and
matter epochs. This is equivalent to a FRW flat universe. For this class $q=1$.
\se{Model with $\eta=\eta_0 H$}
Using eq.(7.13) in eq.(7.14) we obtain
\be
\fr{H'}{H^3}+\fr{3\gamma(1-\bt)}{2R}\ \
\fr{1}{H^2}=\fr{9(1-\bt)\eta_0}{2A}R^{3\gamma-1}
\ee
Let $y=1/H^2$. This becomes
\be
\fr{dy}{dR}-\fr{3\gamma(1-\bt)}{R}y=\fr{-9(1-\bt)\eta_0}{A}R^{3\gamma-1},
\ee
\be
\fr{d}{dR}yR^{-2a}=\fr{-9(1-\bt)\eta_0}{A}R^{3\gamma-1-2a},
\ee
\be
y=\fr{9(1-\bt)\eta_0}{A(2a-3\gamma)}R^{3\gamma},
\ee
and hence
\be
H^2=NR^{-3\gamma}, \ \ N=\fr{A(2a-3\gamma)}{9(1-\bt)\eta_0}.
\ee
Substituting this in eq.(7.11) and using eq.(7.13) yields
$$G=\fr{3N}{8\pi A}=\rm\ const. .$$
Hence
$$ R(t)=N^{\fr{1}{3\gamma}}t^{\fr{2}{3\gamma}}.$$
This reduces to the flat FRW model with constant $G$. This is equivalent to
the solution with $n=1/2$. Therefore the assumption  $\eta\sim H$ is
equivalent to $\eta\sim\rho^{1/2}$ \ [12].

\se{Model with $n=2/3$, \ \ \ $\gamma=2$}
The scale factor is given by
$$R(t)=Ft^{1/2},$$
and
$$\rho(t)=A't^{-3} \ \ ,$$
$$G(t)=Bt \ \ ,$$
$$\eta(t)=\eta_0t^{-2},$$
where F, $A'$ and B are constants.\\
This result is obtainable from Berman [2] if we let $A=\fr{1}{16\pi},\
B=\fr{1}{4}$ and $m=2$. For this model $q=1$
\se{Model with $n=1/2,\ \gamma=2$}
The scale factor is given by
$$R=Ft^{1/3}$$
and
$$\rho=A't^{-2},$$
$$G=B=\rm \ const.,$$
where F and $A'$ are constants.\\
This is the solution for the BD theory for the present phase, as shown by
Berman and Som (1990). For this model $q=2$. This solution also found by
Beesham for Bianchi type I models for $n=0$ (where $\gamma=n+2$, i.e. $\rho\sim t^{-\gamma}$).
Barrow showed that  $\rho\sim t^{-\gamma}$ dominates the viscous
term for all fluids with $1\leq\gamma\leq 2$ \ [3].
\se{Model with $n=1/2 , \ \ \gamma = 2/3 $}
The scale factor is given by
$$R=Ft\ \ ,$$
and
$$\rho=A't^{-2} \ \ ,$$
$$G=\rm \ const.,$$
where F and $A'$ are constants.\\
These are the solutions obtained by Pimentel \ [14] for the scalar field of
the second-self creation theory proposed by Barber, assuming a power law of
the scalar field and the expansion factor. The resemblance is evident if we put
$$n=\fr{n_{\rm P}+3(\gamma-1)}{6\gamma}$$
$n_P : n$ due to Pimentel. When $n=1/2$, the present case, $n_P=3$. For this
class of model $q=0$. There is no horizon problem associated with this solution.\\
\se{Model with n=1/2  \ \ \ $\gamma= 1/3$}
The scale factor is given by
$$R=F t^2,$$
and $$\rho=A' t^{-2}, $$
 $$G=\rm const.,$$
where F and $A'$ are constants.\\
This is the wall-like matter. For this solution $q=-1/2$. This solution
has been obtained by Berman for the radiation universe, i.e. a wall-like
matter behaves the same as radiation in a viscous universe.\\
\se{Model with $n=0 ,\ \  \gamma=1/3$}
 The scale factor is given by
$$R=Ft,$$
and
$$\rho=A't^{-1},$$
$$G=Bt^{-1},$$
where F, $A'$ and B are constants.\\
This solution which solves for the power law is also Machian i.e.
$G\rho\sim H^2$, ( see [25]). Such solution has been noted by Berman and Som for
the constant deceleration type with $m\neq 0 $\ [25].
This solution corresponds to the case $m=1$. In this case we see that the
viscosity is constant, i.e. $\eta=\eta_0$. This solution is a wall-like matter
solution. This model is free of the horizon problem. This solution has been
obtained by Pimentel \ [24] for the solution of the Brans-Dicke theory with a
constant bulk viscosity for $k\neq 0$ solution. He has shown that these
solutions satisfy the Machian condition and the second Dirac hypothesis.
Singh and Devi \ [23] studied cosmological solutions in Brans-Dicke theory
involving particle creation and obtained a similar solution for $k=0$.
Some other solutions are
\se{The Pimentel solution for the scalar tetradic theory A}
\underline{ Case I}\\
This solution of Pimentel [4] is equivalent to our solution provided
$$n_P=(\beta_P-2+3\gamma)$$
and $\beta_P=2-6\gamma(1-n)$,
where the subscript ``P" is the Pimentel value. Therefore this solution is a
viscous solution.
The viscosity ($\eta$) varies as $t^{-\fr{(2n_P-\beta_P+2)}{(2-\beta_P)}}$. The
condition $\beta_P\neq 2$ is equivalent to $n\neq 1$. Note that
$$\gamma=2\left(1-\fr{1}{\om n_P}\right)$$
or $$\om=\fr{2}{3\gamma(2n-1)(\gamma-2)}$$
\underline{ Case II}          \\
This solution is equivalent to our solution provided we make the following
substitution
$$3\gamma(1-n)=1$$
or $$n_{\rm P}=\fr{2n-1}{1-n}=3\gamma-2$$
and therefore
$$\om=\fr{2(n-1)}{(2n-1)(\gamma-2)}.$$
The viscosity coefficient in this case varies as
$$\sim t^{-(1+n_P)} (\beta_P=0).$$
The viscosity term in this case $\sim\exp -(\fr{n_P+1}{n_P+2})t$.
\se{Berman solution}

 Berman studied a constant deceleration model [2]. eqs. (7.14) and (7.15) of Berman are
equivalent to eq.(7.24) provided
$$\beta=2n-1 \leq 0, $$
$$A=\fr{1}{12\pi\gamma^2(1-n)},$$
and
$$B=\fr{2n-1}{3\gamma^2(1-n)^2}.$$
We conclude that Berman solution is equivalent to a  bulk viscous model
with variable $G$ and $\Lambda$.
The viscosity term here varies as $\eta \sim t^{-(1+\fr{B}{4\pi A})} $.\\
 More recently Johri and Desikan (1994) have considered cosmological models in
 Brans-Dicke theory with constant deceleration parameter. Their solution for
 a flat universe,[their eqs.(65)--(67)] are equivalent to our solution,
i.e. eqs.(7.23)--(7.25) for the replacement of\\
$\beta= 3\gamma-3\gamma n-1$ and $\alpha=3\gamma(1-2n).$

\se{Cosmological expansion in the presence of quadratic bulk viscosity $(\zeta)$}
This term appears as $3\zeta H^2$ in the pressure term.\\ Let us  consider $\zeta=\rm const.$
case.
It follows that
$$(9\eta H+3\zeta H^2)\fr{\dot{R}}{R}=\fr{\dot{G}}{G}\rho+ \fr{\dot{\Lambda}}{8\pi G}$$
Using eq.(7.4) we obtain
$$ H'+\fr{3\gamma(1-\bt)}{2R}H-\fr{9(1-\bt)\eta_0}{2}R^{3\gamma-3\gamma n-1}H^2 -\fr{3\zeta(1-\bt)}{2}R^{3\gamma-1}H^3=0$$
This equation admits a power law solution of the form
$$H=\alpha R^m, \ \ \alpha \rm \ \ const.$$\\
Substituting this in the above equation ,\\
we get $m=-\fr{3\gamma}{2} $ and $n=1/2$ \\
Hence
$$ H=\alpha R^{-\fr{3\gamma}{2}},$$
or
$$R=(\fr{3\gamma \alpha}{2})^{\fr{2}{3\gamma}} t^{\fr{2}{3\gamma}}$$
This is the familiar FRW flat universe solution.
If we take a general power law for $\zeta$, i.e. $\zeta \sim \rho^r$ for some
 $r$, it follows that only $r=1$ is possible. This case has been studied by
Wolf \ [5].
He showed that a constant $\rho$ leads to the inflationary solution. This model is
similar to the one considered before ($\eta =\eta_0 \rh$).
We see for all these models one has the Machian solution $G\rh\sim H^2$.
\se{Brans-Dicke solution}
In the Brans-Dicke theory [29] the scalar field is related to the gravitational
constant $G$ as $\phi\propto 1/G$.
This theory is equivalent to a bulk viscous solution with
\be
n=\fr{3\om+2}{6(1+\om)}\ ,
\ee
in which case the viscosity coefficient varies as
\be
\eta\sim t^{-\fr{3\om+2}{3\om+4}}.
\ee
As in the Brans-Dicke theory when $\om\rightarrow\infty$ the theory approximates
to Einstein de Sitter, in the present case we have $n\rightarrow 1/2 $.
\se{References}
$[1]$ M.Novello and R. A. Araujo, {\bf Phys. Rev.D22}, 260(1980)\\
$[2]$ M.S.Berman, {\it Gen. Rel. Gravit. 23}, 465(1991)\\
$[3]$ J.D.Barrow, {\it Nuc.Phys.B310}, 743(1987)\\
$[4]$ L.O. Pimentel, {\bf Astrophys. Space Sci.}, 387(1987) \\
$[5]$ C.Wolf, {\it S.-Afr.Tydskr Fis.14}(1991)   \\
$[6]$ A.Beesham , {\it Gen. Rel. Gravit.26}, 159, (1994) \\
$[7]$ S.D. Maharaj and R.Naidoo, {\it Astrophys. Space Sci.208}.261(1991)  \\
$[8]$ R.F. Sistero ,{\it Gen. Rel. Gravit. 23}, 1265 (1991) \\
$[9]$ M.S.Berman , {\it Nouvo Cimento 74B}, 182(1983)  \\
$[10]$ M.S.Berman, {\it IJTP 29} 571(1990)   \\
$[11]$ A.-M.M. Abdel Rahman, {\it Gen. Rel. Gravit. 22}, 655(1990)   \\
$[12]$ O.Gron, {\bf Astrophys. Space Sci.173},191(1990)  \\
$[13]$ G. Murphy, {\it Phys. Rev. D8}, 4231(1973)    \\
$[14]$ L.O. Pimentel, {\it Astrophys. Space Sci. 116}, 395(1985)\\
$[15]$ M.S. Berman and M.M.Som, {\it IJTP 29}, 1411(1990)    \\
$[16]$ C. Brans and R.H. Dicke, {\it Phy. Rev. D124}, 203(1961)  \\
$[17]$ M. \"{O}zer and M.O.Taha , {\it Nuc. Phy. B287}776(1987)    \\
$[18]$ Z. Pazamata, {\it IJTP 31}, 2115(1987)      \\
$[19]$ D. Kalligas, P. Wesson, and C.W.Everitt, {\it Gen. Rel. Gravit. 24}, 351(1992) \\
$[20]$ A.Beesham, {\it  Phy. Rev. D48}, 3539(1993)   \\
$[21]$ Z.Golda, M.Heller, and M.Szydlowski, {\it Astrophys. Space.Sci. 90}, 313(1983) \\
$[22]$ Z. Klimek, {\it Nouvo Cimento, 35B} 249(1976)      \\
$[23]$ R.K.Singh and A.R. Devi, {\it  Astrophys. Space Sci.155}, 233(1989)    \\
$[24]$ l.O. Pimentel, {\it IJTP 33}, 1335(1994)    \\
$[25]$ M.S.Berman, {\it IJTP 29}, 571(1990)    \\
$[26]$ S. Berman and S.S. Som, {\it Phys.Lett.139A}, 119(1989)\\
$[27]$ V.B Johri and K. Desikan, {\it Gen. Rel. Gravit.26}, 1217(1994)\\
$[28]$ T.Padmanabhan and S.M. Chitre, {\it Phys. Lett. A120},433(1987)\\
$[29]$ C. Brans and R.H. Dicke, {\it Phy. Rev.D124}, 125 (1962)\\
$[30]$ J. C. Carvalho, J. A. S. Lima and I. Waga, Phys. Rev. D. 46 (1992) 2404
\chapter{A flat Viscous Universe with Increasing $G$}
\se{ Introduction} In most variable $G$ models [7,8]  $G$ is a
decreasing function of time. But the possibility of increasing $G$
has only been considered recently [2]. Massa $[5]$ proposed a
model in scale of increasing $G$ depending on a ``maximal power
hypothesis (MPH)." Recently [1], we have considered a cosmological
model with variable $G$ and
 $\Lambda$ and bulk viscosity. Various solutions are listed and all of
 them satisfy Mach's condition [4].
\se{ The Model}
In a Robertson Walker universe
\be
d\tau^{2}= dt^{2}-
R^{2}(t)[\fr{dr^{2}}{1-kr^2}+r^{2}(d\theta^{2}+\sin\theta^{2}d\phi^{2})]
\ee
where $k$ is the curvature index.\\
Einstein's field equations with time dependent cosmological and gravitational
``constants''
\be R_{\mu\nu}-\fr{1}{2}g_{\mu\nu}R=8\pi GT_{\mu\nu}+ \Lambda g_{\mu\nu}
\ee
and the perfect fluid energy momentum tensor
\be T_{\mu\nu}=(\rho+p)u_{\mu}u_{\nu}-pg_{\mu\nu}
\ee
yield the two independent equations
\be
3\fr{\ddot{R}}{R}= -4\pi G(3p+\rho-\fr{\Lambda}{4\pi G}),
\ee
\be
3\fr{\dot{R}^{2}}{R^2}=8\pi G(\rho+\fr{\Lambda}{8\pi G})-\fr{3k}{R^2}.
\ee
Elimination of $\ddot{R}$ gives
\be
3(p+\rho)\dot{R}=-(\fr{\dot{G}}{G}\rho+\dot{\rho}+\fr{\dot{\Lambda}}{8\pi G})R.
\ee
The conservation of energy and momentum yields
\be
3(p+\rho)=-R\fr{d\rho}{dR}.
\ee
The effect of bulk viscosity in the field equation is to replace
$p$ by $p-3\eta H$, where $\eta$ is the viscosity coefficient. It follows
immediately that
\be
9\eta H\dot{R}=(\fr{\dot{G}}{G}\rho+\fr{\dot{\Lambda}}{8\pi G})R
\ee
and
\be
\dot{\rho}+3H(\rho+p)=0    .
\ee
Equation (8.8) can be written as
\be
9\eta \fr{H}{R}=\fr{G'}{G}\rh+\fr{\Lambda'}{8\pi G},
\ee
where prime denotes derivative w.r.t. scale factor $R$ and dot is the derivative
w.r.t. cosmic time $t$. In what follow we will consider the flat universe, $k=0$.
Equation (8.5) becomes
\be
3H^2=8\pi G\rh+\Lambda.
\ee
We take the Chen and Wu ansatz for $\Lambda$ $[3]$
\be
\Lambda=\fr{3\alf}{R^2} ,\ \ \ \alf \ \ \rm const.,
\ee
and the viscosity to have the form
\be
\eta=\eta_0\rh^n, \ \ \ \ \eta_0\ge 0, n\ \rm const..
\ee
Using the equation of the state $p=(\gamma-1)\rh$ in eq.(8.9) we obtain
\be
\rh=AR^{-3\gamma}, \ \  \ A\ \ \rm const..
\ee
Substituting eqs.(8.12), (8.13) and (8.14) in (8.10) yields
\be
G'-9\eta_0A^{n-1}R^{-3\gamma n+3\gamma-1} H G-\fr{3\alf}{4\pi A}R^{3\gamma-3}=0.
\ee
This equation admit a power law solution of the form
\be
G=FR^m, \ \ \ \ F, m \ \ \ \rm const..
\ee
Inserting this in the above equation we get
\be
H=C\ R^{3\gamma(n-1)}
\ee
and the condition,
\be
m=3\gamma-2,
\ee
where $ C=(m-\fr{3\alf}{4\pi FA})/(9\eta_0A^{n-1})$.
From eqs.(8.16) and (8.18) we see that in the radiation epoch ($\gamma=4/3$)
$G\propto R^2$ and in the matter epoch $G\propto R$. As long as $\gamma > 2/3, G$
continues to increase.

This result agrees with that obtained by Abdel Rahman $[2]$ for the critical
density model. Hence his model is equivalent to a viscous model.
Now eq.(8.17) gives
\be
R=[3\gamma(1-n)C]^{[1/3\gamma(1-n)]}t^{[1/3\gamma(1-n)]}
\ee
Hence eqs.(8.16), (8.14) and (8.12) become
\be
G=G_0 t^{[(3\gamma-2)/3\gamma(1-n)]},
\ee
\be
\rh=\rh_0t^{[-1/(1-n)]},
\ee
\be
\Lambda=\Lambda_0t^{[-2/3\gamma(1-n)]}\ \ ,
\ee
and
\be
H=\fr{1}{3\gamma(1-n)t}\ \ ,
\ee
where $G_0, \rh_0 $ and $\Lambda_0$ are constants.
For an expanding Universe, i.e. $H>0$, we must have $C > 0$. This implies  $\alf < \fr{4}{3}\pi FA(3\gamma-2)$, and for $\alf > 0$, $\gamma>2/3$.
Since some authors think that the variation of $\Lambda \propto t^{-2}$ is essential,
we impose the condition
\be
3\gamma(1-n)=1.
\ee
Hence eqs.(8.19) and (8.20) become
\be
R=R_0t,     \ \ \ R_0\ \ \rm const.
\ee
\be
G=G_0t^{[(2n-1)/(1-n)]},  \ \ G_0\ \ \rm const..
\ee
This solution has been obtained by Ref.1 for a different ansatz of $\Lambda$.
In the next section we will consider two cases separately.
\se{I. Radiation dominated (RD) universe}
This is characterized by $p=\fr{1}{3}\rho$.
The condition (8.24) gives $n=3/4$. Hence, eqs.(8.21), (8.20) and (8.13)  become
\be
\rho\propto t^{-4}       \ ,
\ee
\be
G\propto t^2             \ ,
\ee
and
\be
\eta\propto t^{-3}          \ .
\ee
We see that $G$ increases with time. In Ref.2, for $t\gg R_0$ the gravitational
constant $G\propto t^2$. Though the model of Ref.2 is nonsingular and closed yet
it evolves towards this singular viscous model. One also notices that  $T\propto R^{-1}$
 a result that is expected to hold in this era.
An increasing $G$ is recently supported by Massa [6].
\se{II. Matter dominated (MD) universe}
This is a dust filled universe ($\gamma=1)$. The condition (8.24) gives $n=2/3$.
Hence, eqs.(8.21), (8.20) and (8.13) become
\be
\rho\propto t^{-3}\ \ \ ,
\ee
\be
G\propto t   \ \ ,
\ee
and
\be
\eta\propto t^{-2} \ .
\ee
We see that $G$ increases linearly with time. This solution has been found by Berman [3] for the Bertolami solution for
the Brans-Dicke theory with a time dependent cosmological constant. Moreover, Ref.2
predicts that $ G \propto R $ for $ R\rightarrow \infty $ (asymptotically).
\se{An inflationary solution}
This corresponds to the case $n=1$. Hence, eq.(8.17) gives
\be
H=C=\rm const. .
\ee
Therefore
\be
R=\rm const.\exp(Ct)
\ee
This solution has been obtained by $[9,10]$.
Hence eqs.(8.14) and (8.15) become
\be
\rh=N\exp(-3\gamma H_0t)\ \ , \ N  \ \rm const
\ee
and
\be
G=M\exp[(3\gamma-2)H_0t], \ \ M \ \ \rm const. .
\ee
\se{Concluding remarks}
We have analyzed a flat viscous cosmological model with varying $G$ and $\Lambda$.
 The gravitational constant is shown to increase quadratically with time in the pure radiation era and linearly in the matter dominated era.
 We have also shown that Abdel Rahman model approaches this model asymptotically, i.e.
for $t\gg R_0$.
 The model assumes to solve the horizon and monopole problems of the standard model.
 In this model we relax the assumption of the critical density.
 The cosmological constant retains its evolution with time, $\Lambda \propto t^{-2}$, since this
 is assumed to be fundamental. We see that the viscosity becomes more important in
 the matter than in the radiation epoch. This model corresponds to one of the models we have considered
 recently.
\newpage
\se{References}
$[1]$ A. I. Arbab, {\it Gen. Rel. Gravit.29,} 61 (1997)\\
$[2]$ A.-M. M. Abdel Rahman, {\it Gen. Rel. Gravit.} 22, 655(1990)   \\
$[3]$ M. S. Berman and M.M.Som, {\it IJTP}, 29, 1411(1990)    \\
$[4]$ W. Chen and Y. S. Wu, {\it Phy. Rev. D41}, 695(1990)\\
$[5]$ M. S. Berman, {\it IJTP}, 29, 571(1990)    \\
$[6]$ C. Massa, {\it Astrophys. and Space. Sci}, 232, 143 (1995)\\
$[7]$ S. Weinberg, Gravitation and Cosmology (1972)(Wiley, NY)\\
$[8]$ E. B. Norman, {\it Am.J.Phys.}, 54, 317 (1986)\\
$[9]$ J. D. Barrow, {\it Nucl. Phys. B310}, 743(1988)\\
$[10]$ S. D. Maharaj and R. Naidoo, {\it Astrophys. Space Sci. 208}, 261 (1993)\\
\newpage
\se{Appendix A}\footnote{See e.g. L.D.Landau and E.M. Lifshitz,
{\it Fluid Mechanics} (Pergamon Books Ltd. 1987)} In order to
obtain the equations describing the motion of a viscous fluid, we
have to include some additional terms in the equation of motion of
an ideal fluid. The equation of motion of a viscous fluid may
therefore be obtained by adding to the ideal momentum flux a term
$\sigma_{ik}'$ which gives the irreversible viscous transfer of
momentum in the fluid. Thus we write the momentum flux density in
a viscous fluid in the form $$T"_{ik}=\rh
v_iv_k-p\delta_{ik}+\sigma_{ik}'.$$ The general form of the tensor
$\sigma'_{ik}$ can be found as follows. Processes of internal
friction occur in a fluid only when different fluid particles move
with different velocities, so that there is a relative motion
between various parts of the fluid. Hence $\sigma'_{ik}$ must
depend on the space derivatives  of the velocity. If the velocity
gradients are small we may suppose that the momentum transfer due
to viscosity depends only on the first derivatives of the
velocity. To the same approximation, $\sigma'_{ik}$ may be
supposed a linear function $\fr{\partial v_i}{\partial x_k}$.There
can be no term in $\sigma'_{ik}$ independent of $\fr{\partial
v_i}{\partial x_k}$, since $\sigma'_{ik}$ must vanish for $v=\rm
constant$. Next, we notice that $\sigma'_{ik}$ must also vanish
when the whole fluid is in uniform rotation, since it is clear
that in such a motion no internal friction occurs in the fluid.
The sum $$\fr{\partial v_i}{\partial x_k}+\fr{\partial
v_k}{\partial x_i}$$ are linear combination of the derivatives
$\fr{\partial v_i}{\partial x_k}$, and vanish when $v=\Omega\times
r$, where $\Omega$ is the angular velocity. Hence $\sigma'_{ik}$
must contain just these symmetrical combinations of the
derivatives $\fr{\partial v_i}{\partial x_k}$. The most general
tensor of rank two satisfying the above condition is
$$\sigma'_{ik}=\zeta(\fr{\partial v_i}{\partial x_k}+\fr{\partial
v_k}{\partial x_i}-\fr{2}{3}\delta_{ik}\fr{\partial
v_\ell}{\partial x_\ell})-\eta\delta_{ik}\fr{\partial
v_\ell}{\partial x_\ell}$$ with coefficients $\zeta$ and $\eta$
independent of the velocity. In making this statement we use the
fact that the fluid is isotropic, as a result of which its
properties must be described by scalar quantities only ( in this
case, $\zeta, \eta$). The constants $\zeta$ and $\eta$ are called
coefficients of viscosity. Note that $\zeta$ and $\eta$ are
functions of temperature and pressure and therefore are not
constants throughout the fluid. \se{Appendix B} The apparent
luminosity is related to the absolute one as by
\be
l=\fr{LR^2(t_1)}{4\pi R^4(t_0)r_1^2}
\ee
where $r_1$ is the source coordinate, and $t_1, t_0$ are the time of emission and
reception respectively. Hence
\be
d_L=\fr{R^2(t_0)}{R^2(t_1)}r_1.
\ee
But
\be
\int_{t_1}^{t_0}\fr{dt}{R(t)}=\int_0^{r_1}\fr{dr}{\sqrt{1-kr^2}}\equiv
f(r_1). \ee The scale factor can be expanded in power of ($t_0-t)$
as
\be
R(t)=R(t-(t_0-t))=R(t_0)[1-(t_0-t)H_0-1/2(t_0-t)^2q_0H_0^2+....]
\ee
where
$f(r_1)=\sin^{-1}r_1,r_1,\sin r_1,$ according to whether $k=1,0,-1$.\\
The red-shift $z$ is
\be
z=(t_0-t_1)H_0+(t_0-t_1)^2(q_0/2+1)H_0^2+....
\ee
Inverting this, the look-back time is
\be
t_0-t_1=H_0^{-1}z-H_0^{-1}(1+q_0/2)z^2+...
\ee
Therefore
$$r_1+O(r^3)=R_0^{-1}[t_0-t_1+1/2H_0(t_0-t_1)^2+...$$
$$=(R_0H_0)^{-1}[z-1/2(1+q_0)z^2+...]$$
\be
d_L=H_0^{-1}[z+1/2(1-q_0)z^2+...]
\ee
Hence
$$\ell=\fr{L}{4\pi d_L^2}=\fr{LH_0^2}{4\pi z^2}[1+(q_0-1)z+...]$$
$\ell$ is usually expressed in terms of apparent bolometric magnitude $m$.
This is defined by
$$l=10^{-2m/5}\times 2.52\times 10^{-5}\ \rm erg/cm^2/s.$$
The absolute luminosity is defined as the apparent magnitude of the source
would have at a distance of $10pc$.
$$L=10^{-2M/5}\times 3.02\times 10^{35}\ \rm erg\ s^{-1}.$$
The distance modulus can be found as follows
$$d_L=10^{1+(m-M)/5}\ pc$$
and hence
$$m-M=25-5\log H_0+1.086(1-q_0)z+...$$
or
$$ m-M=5\log d_L-5 $$
\end{document}